\documentclass[11pt]{article}
\usepackage{epsfig,amsmath,amssymb,amsthm,amsopn}
\usepackage[footnotesize,bf]{caption}
\newcommand{\be}{\begin{equation}}
\newcommand{\ee}{\end{equation}}
\newcommand{\bea}{\begin{eqnarray}}
\newcommand{\eea}{\end{eqnarray}}
\newcommand{\nn}{\nonumber}
\newcommand{\half}{{\scriptstyle \frac{1}{2}}}
\begin{document}
\setlength{\captionmargin}{20 pt}
\title{Asymmetric Chern-Simons number diffusion from CP-violation}
\author{Bert-Jan Nauta and Alejandro Arrizabalaga\footnote{nauta@science.uva.nl, arrizaba@science.uva.nl}\\
\\
 Institute for Theoretical Physics\\
Valckenierstr. 65, 1018 XE Amsterdam}
\date{February 12, 2002}
 \maketitle
\abstract{We study Chern-Simons number diffusion in a
SU($2$)-Higgs model with CP-odd dimension-eight operators. We find
that the thermal average of the magnitude of the velocity of the
Chern-Simons number depends on the direction of the velocity. This
implies that the distribution function of the Chern-Simons number
will develop an asymmetry. It is argued that this asymmetry
manifests itself through a linear growth of the expectation value
of the third power of the Chern-Simons number. This linear
behavior of the third power of a coordinate of a periodic
direction is verified by a numerical solution of a one-dimensional
Langevin equation. Further, we make some general remarks on
thermal averages and on the possibility of the generation of the
baryon asymmetry in a non-equilibrium situation due to asymmetric
diffusion of the Chern-Simons number.}

\section{Introduction}
An important cosmological observation is the matter-antimatter
asymmetry in the present universe. This asymmetry may be
quantified by the baryon-to-photon ratio, whose observational
value is \cite{olive} \begin{equation}
\frac{n_{B}}{n_{\gamma}}=(1.55-4.45)\times 10^{-10},
\label{obsasymmetry} \end{equation} with \(n_{B}\) the baryon
number density and \(n_{\gamma}\) the photon density. An
interesting question is how this asymmetry was generated in the
early universe.
In his 1967 paper Sakharov \cite{sakharov} notes that baryon number can only be generated when\\
1) baryon number is not conserved,\\
2) the transformations C and CP are not symmetries,\\
3) there is a departure from equilibrium.

In 1976 't Hooft \cite{hooft} showed that in the electroweak
theory the first requirement is satisfied. The non-trivial vacuum
structure of the SU(\(2\))-gauge theory in combination with the
anomaly equation implies that a change in Chern-Simons number,
\(N_{\rm CS}\), is accompanied by a change in baryon number,
\(B\),
\bea
B(t)-B(0)&=&3\left[N_{\rm CS}(t)-N_{\rm CS}(0)\right]\nn\\
&=&\frac{3g^2}{32\pi^2} \int_{0}^{t} dt' \int d^3x
F_{\mu\nu}^{a}\tilde{F}^{\mu\nu a}, \label{relCSbar} \eea with
\(g\) the SU(\(2\))-gauge-coupling, \(F_{\mu\nu}^{a}\) the field
strength, and \(\tilde{F}^{\mu\nu a}\) its dual. At zero
temperature changes in the Chern-Simons number involve instanton
processes. The rate of these processes is very small \cite{hooft}.
However, at high temperature the rate is much larger. As has long
been recognized this opens up the possibility that the baryon
asymmetry was generated at temperatures of about \(100\) GeV
\cite{kuzmin}, when the (elementary) particles that made up the
plasma may be described by the electroweak theory.

However, within the standard model the amount of CP-violation is
probably too small to account for the observed asymmetry
(\ref{obsasymmetry}) \cite{rubakovshaposhnikov}. One way to deal
with this is to include effective non-renormalizable operators
that parameterize the CP-violation of a more fundamental theory.
The lowest-dimensional operator in a SU(\(2\))-Higgs theory is
\(\phi^{\dagger}\phi F_{\mu\nu}^{a}\tilde{F}^{\mu\nu a}\), with
\(\phi\) the Higgs field. It may be included in the effective
action as \begin{equation}S^{[6]}=\frac{3g^2 \delta_{\rm
CP}^{0}}{32\pi^2 M^2}\int d^4x \phi^{\dagger}\phi
F_{\mu\nu}^{a}\tilde{F}^{\mu\nu a}, \label{dimsix}\end{equation}
with \(M\) a mass and \(\delta_{\rm CP}^{0}\) a coefficient that
ideally may be derived from the fundamental theory. In the current
investigation dimension-eight operators will play a more important
role. They may be included in the action as
\begin{equation}S^{[8]}=\frac{3g^2}{32\pi^2 M^4}\int d^4 x
\left[ \delta_{\rm CP}^{1}(D_{\rho} \phi)^{\dagger} (D^{\rho}
\phi) -\delta_{\rm CP}^2 F_{\rho\sigma}^{b}F^{\rho\sigma b}
+\delta_{\rm CP}^3(\phi^{\dagger}\phi)^2 \right]F_{\mu\nu}^a
\tilde{F}^{\mu\nu a}, \label{dimeight}\end{equation} with
\(D_{\rho}\) the covariant derivative.

In this paper we study the combination of the first two of
Sakharov's requirements, although we were unable to refrain from
making some remarks on the inclusion of the third one in our
study. That is, we study the motion of the Chern-Simons number in
the SU(\(2\))-Higgs action extended with (\ref{dimsix}) and
(\ref{dimeight}). It is well known that in a pure SU($2$)-Higgs
theory (without extra CP-odd operators and without fermions) the
long-time behavior of the Chern-Simons number may be viewed as a
diffusion process in one dimension, that can be specified by the
expectation values \bea
\langle N_{\rm CS}(t)-N_{\rm CS}(0)\rangle&=&0,\label{diffonepow}\\
\langle\left[N_{\rm CS}(t)-N_{\rm
CS}(0)\right]^2\rangle&=&2V\Gamma_{\rm sph}t,
\label{difftwopow}\eea with \(V\) the volume and \(\Gamma_{\rm
sph}\) the sphaleron rate. The sphaleron rate plays an important
role in scenarios for electroweak baryogenesis, since it
determines the rate of the baryon-number violating processes. Much
effort has gone into the determination of the sphaleron rate, both
in the symmetric phase
\cite{asy},\cite{bodeker},\cite{bodekermoorerummukainen} as well
as in the broken phase
\cite{arnoldmclerran},\cite{khlebnikov},\cite{moore}. Here, only
the broken-phase sphaleron-rate will be needed
\begin{equation}\Gamma_{\rm
sph}=\kappa T^4 e^{-\beta E_{\rm sph}},
\label{sphrate}\end{equation} with the sphaleron energy \(E_{\rm
sph}\), and \(\kappa\) a dimensionless coefficient that depends on
the temperature, the Higgs mass and expectation value, and the
gauge coupling. The physical picture that underlies
(\ref{sphrate}) is that of classical transitions from one
(classical) vacuum to the next over a potential barrier of height
\(E_{\rm sph}\). On the basis of this picture, the gauge-Higgs
dynamics will be treated classically in this paper.

The idea that the inclusion of CP-violating operators may have an
interesting effect on the diffusion of the Chern-Simons number
stems from the fact that the Chern-Simons number itself is CP-odd.
Therefore the motion over the barrier towards positive
Chern-Simons numbers may differ from the motion in the direction
of negative Chern-Simons numbers.  It is clear then, that the
distribution function of the Chern-Simons number does not need to
develop symmetrically. Or put differently: there is no symmetry
argument why expectation values of odd powers of \(N_{\rm
CS}(t)-N_{\rm CS}(0)\) should vanish. In fact, we will show that
an asymmetry in the distribution function will develop indeed (as
may have been inferred from the title). That is, an initially
symmetric distribution function will develop an asymmetry. In the
presence of the CP-odd operators in (\ref{dimeight}) an equation
for the expectation value of the third power of \(N_{\rm
CS}(t)-N_{\rm CS}(0)\) needs to be added to equations
(\ref{diffonepow}) and (\ref{difftwopow}) to characterize the
diffusion. We will argue that this expectation value increases
linearly in time in the broken phase:
\begin{equation}
\langle\left[N_{\rm CS}(t)-N_{\rm CS}(0)\right]^3\rangle \sim t,
\end{equation}
A more detailed conjecture is given in equation (\ref{conjCS}).

We end the introduction with a short outline of the rest of the
paper. In the next section the dynamics of the gauge-Higgs system
will be projected on a one-dimensional path through the
configuration space. This will simplify the analysis of the rest
of the paper. In section \ref{sec:avvel} we will calculate the
thermal average of the velocity when the system moves in the
direction of positive or negative Chern-Simons numbers. In section
\ref{sec:conj} a conjecture for the time dependence of \(\langle
[N_{\rm CS}(t)-N_{\rm CS}(0)]^3\rangle\) is presented, where it is
argued to grow linearly in time. This linear growth is also
present in a simple random walk model as is shown in appendix
\ref{sec:RWM}. Section \ref{sec:remarks} contains some general
remarks on statistical averages. In the following section, we
derive a simple stochastic equation that is useful to verify some
of the analytic results. Also some numerical solutions are
presented. In section \ref{sec:bar} we discuss how asymmetric
diffusion of the Chern-Simons number may yield a non-zero baryon
number in a out-of-equilibrium situation. Also some comments are
made on the constraints on such a model. We end with a summary in
section \ref{sec:concl}.

\section{Projection of the dynamics of the gauge and Higgs fields
 onto one special direction in configuration space}\label{to1d}

To simplify the later study of the effect of the CP-violating
operators in (\ref{dimsix}) and (\ref{dimeight}) on the diffusion
of the Chern-Simons number, the gauge-Higgs-field dynamics is
projected on a single path through the configuration space. In
order to include sphaleron transitions, we use a path that runs
from a classical vacuum to the sphaleron configuration and further
to the next vacuum. Manton \cite{manton} defined such a path in
his proof that a non-contractible loop exists in configuration
space of the SU($2$)-Higgs model. This is the path that will be
used in the following. The path of Manton is not the
minimal-energy path, which was constructed in \cite{akiba}. But
the precise path will not be important for the following rather
general arguments and we expect that the results will suffice as
order of magnitude estimates.

We parameterize the path of Manton by the time-dependent
coordinate \(\Theta\). The fields are given in the radial gauge
(\(\partial_{i} A^{i a}=0\)) by
 \bea \mbox{gauge}&&
A_{\mu}^{a}\sigma^{a}=
\frac{-2i}{g}f(\rho)[\partial_{\mu}U(\Theta)]U^{-1}(\Theta),
\label{gpathparametrization}\\
\mbox{Higgs}&& \phi=\frac{1}{2}\sqrt{2} v \left\{h(\rho) U(\Theta)
\left(\begin{array}{c} 0 \\ 1\end{array} \right)-
i[1-h(\rho)]\cos\Theta\left(\begin{array}{c} 0 \\ 1\end{array}
\right)\right\}, \label{hpathparametrization}\eea with \(v\) the
Higgs expectation value and \(\rho=gvr\), where \(r=|\vec{x}|\).
The \(\Theta\)-dependent SU($2$)-matrix is given by
\begin{equation}
U(\Theta)=\frac{1}{r}\left(\begin{array}{cc} z & x+iy \\ -x+iy & z
\end{array} \right)\sin\Theta +\left(\begin{array}{cc} i & 0 \\ 0
& -i \end{array} \right)\cos\Theta .
\label{fieldpar}\end{equation} The functions \(f\) and \(h\)
satisfy the boundary conditions \bea
f\rightarrow 0 && h\rightarrow 0 \;\;\;\;\;\;\; r\rightarrow 0,\nn\\
f\rightarrow 1 && h\rightarrow 1 \;\;\;\;\;\;\; r\rightarrow
\infty . \label{boundcond}\eea

The form of the fields (\ref{gpathparametrization}),
(\ref{hpathparametrization}) is a non-static generalization of the
fields considered in \cite{manton}. When \(\Theta\) is chosen to
be time-independent the fields given by
(\ref{gpathparametrization}), (\ref{hpathparametrization}) can be
identified (up to a global U($1$) gauge transformation) with the
fields ($3.17$) in \cite{manton} (the coordinate \(\Theta\)
corresponds to \(\mu\) in \cite{manton}).

For the functions \(f\) and \(h\) the Ansatz b of Klinkhamer and
Manton \cite{mantonklinkhamer} will be used \bea f(\rho)&=&\left\{
\begin{array}{llr}
\frac{\rho^2}{A(A+4)} & \rho\leq A\\
1-\frac{4}{A+4}\exp[\frac{1}{2}(A-\rho)] &\rho\geq A
\end{array}\right.,\label{gpar}\\
h(\rho)&=&\left\{
\begin{array}{llr}
\frac{\sigma B+1}{B(\sigma B+2)}\rho & \rho \leq B\\
1-\frac{B}{\sigma B+2}\frac{1}{\rho}\exp[\frac{1}{2}(B-\rho)]
&\rho\geq B
\end{array}\right. ,
\label{hpar}\eea with \(\sigma=(\lambda/2g^2)^{\half}\), where
\(\lambda\) is the Higgs self-coupling constant. The parameters
\(A,B\) were determined by minimizing the energy for the static
field configuration at \(\Theta=\pi/2\) in
\cite{mantonklinkhamer}, in order to obtain an estimate for the
sphaleron energy.  In this way the parameters depend only on
\(\sigma\). We take \(\sigma = 1\), for which the parameters have
the values \(A=1.15\) and \(B=1.25\) \cite{mantonklinkhamer}.

Now that the dynamics has been restricted to the path described by
(\ref{gpathparametrization}) and (\ref{hpathparametrization}) we
may rewrite the SU($2$)-Higgs action, $S$, in terms of the
coordinate \(\Theta\)
\begin{equation}S[\dot{\Theta},\Theta]=\frac{4\pi
v}{g}\int dt \left[\left(a_1+a_2\sin^2 \Theta\right)
\frac{\dot{\Theta}^2}{(gv)^2} -\left( a_3\sin^2\Theta + a_4
\sin^4\Theta\right)\right]. \label{actiontheta} \end{equation} The
CP-odd action (\ref{dimeight}) in terms of \(\Theta\) reads
\begin{equation}
S^{[8]}[\dot{\Theta},\Theta]= \frac{4\pi v^2}{M^4}\int dt \left[
b_1 \delta_{\rm CP}^{1}+ b_2 \delta_{\rm CP}^2 + (b_3\delta^1_{\rm
CP}+b_4\delta^2_{\rm CP})\sin^2\Theta\right] \dot{\Theta}^3
\sin^2\Theta, \label{CPactionstheta}\end{equation} where total
time-derivatives have been neglected.

The action \(S^{[6]}\) (\ref{dimsix}), that includes the
dimension-six operator \(\phi^{\dagger}\phi F\tilde{F}\), gives
only a total time-derivative, just as the operator
\((\phi^{\dagger}\phi)^2F_{\mu\nu}^{a}\tilde{F}^{\mu\nu a}\) in
(\ref{dimeight}). Therefore the inclusion of \(S^{[6]}\) will not
affect the motion along the path parameterized above. When the
full field-dynamics is considered, the dimension-six operator may
have an effect on the motion along the Chern-Simons number
direction in a non-equilibrium situation, see e.g.
\cite{garciabellido}. In equilibrium however, the system
oscillates around one vacuum or, during a transition, around the
 minimal-energy path that connects two classical vacua (the path
 that we have approximated by (\ref{gpathparametrization}) and (\ref{hpathparametrization})).
  Since, the oscillations average out
 in equilibrium, we do not expect the inclusion of \(S^{[6]}\) to
affect the diffusion of the Chern-Simons number.

The coefficients \( a_1\), \(a_2\), \(a_3\), \(a_4\), \(b_1\),
\(b_2\), \(b_3\) \(b_4\) are given by the integrals
\begin{eqnarray}
a_1&=&\int_{0}^{\infty} d\rho \,\rho^2\left[(\partial_{\rho}f)^2
+\frac{1}{2}(h-f)^2
\right]=2.03,\\
a_2&=&\int_{0}^{\infty} d\rho \,\left[8f^2(1-f)^2
+\frac{1}{2}(1-h^2)(1-f^2)\rho^2
\right]=2.53, \\
a_3&=&\int_{0}^{\infty} d\rho
\left[4(\partial_{\rho}f)^2+\frac{1}{2}\rho^2(\partial_{\rho}h)^2+h^2(1-f)^2\right.\nn \\
& &\hspace{1cm}\left.-2fh(1-f)(1-h)+f^2(1-h)^2\right]=1.41, \\
a_4&=&\int_{0}^{\infty} d\rho \left[8f^2(1-f)^2/\rho^2+2fh(1-f)(1-h)\phantom{\frac{1}{4}}\right.\nn \\
& &\hspace{1cm}\left.-f^2(1-h)^2+\frac{1}{4}\rho^2(1-h^2)^2\right]
=0.70, 
\end{eqnarray}
\begin{eqnarray}
b_1&=&\frac{9}{2} \int_{0}^{\infty} d\rho\,(h-f)^2
(\partial_{\rho}f) f (1-f)=0.09, \\
b_2&=&9 \int_{0}^{\infty} d\rho (\partial_{\rho}f)^3 f (1-f)=0.10, \\
b_3&=&\frac{9}{2}\int_{0}^{\infty} d\rho
(1-h^2)(\partial_{\rho}f)
f(1-f)(1-f^2)=0.19,\\
b_4&=&72 \int_{0}^{\infty} d\rho
\frac{1}{\rho^2}(\partial_{\rho}f) f^3(1-f)^3 =0.16.
\label{coefficients}
\end{eqnarray}

Let us shortly discuss the actions (\ref{actiontheta}) and
(\ref{CPactionstheta}). In the action (\ref{actiontheta}) we
recognize the potential barrier \cite{manton},
 \begin{equation}
V(\Theta)=\frac{4\pi v}{g}\left[a_3 \sin^2\Theta + a_4
\sin^4\Theta\right], \label{potentialbarrier}\end{equation}
between different vacua. The sphaleron energy reads
\begin{equation}E_{\rm sph}=V(\Theta=\frac{1}{2}\pi)=
(a_3+a_4)\frac{4\pi v}{g}= 2.11 \frac{4\pi v}{g}. \end{equation}
The sphaleron energy determines the exponential suppression of the
sphaleron rate (\ref{sphrate}). Another quantity that enters the
sphaleron rate is (the real part of) the frequency of the negative
mode, \(\omega_{-}\), at the sphaleron configuration
\cite{arnoldmclerran}. From the action (\ref{actiontheta}) it may
be determined as \begin{equation}
\omega_{-}^2=\frac{a_3+2a_4}{a_1+a_2}(gv)^2\approx 0.61 (gv)^2.
\end{equation} In agreement with the order of magnitude estimate
\(\omega_{-}\sim gv\) in \cite{arnoldmclerran}.

The action (\ref{CPactionstheta}) is odd in \(\Theta\). This can
be understood by realizing that a CP-transformation in terms of
\(\Theta\) is \begin{equation}\Theta\rightarrow -\Theta,
\label{CPtranTheta}\end{equation} since the action
(\ref{dimeight}) is CP-odd, so should (\ref{CPactionstheta}).

Now that the SU($2$)-Higgs action including the effective CP-odd
operators has been immensely simplified, namely to the
one-dimensional action (\ref{actiontheta}) $+$
(\ref{CPactionstheta}), the question: "does the Chern-Simons
number diffuse asymmetrically under the influence of the
dimension-eight operators in the action (\ref{dimeight})?" may be
cast in the form: "does the factor \(\dot{\Theta}^3\) in
(\ref{CPactionstheta}) lead to an asymmetric diffusion  of
\(\Theta\)?". What is meant by "diffusion of \(\Theta\)" may
require some explanation. From equations
(\ref{gpathparametrization}),(\ref{hpathparametrization}), and
(\ref{fieldpar}), one sees that \(\Theta =0\) and \(\Theta=2\pi\)
correspond to the same (static) gauge and Higgs field
configuration. Even \(\Theta=0\) and \(\Theta=\pi\) correspond to
the same physical state, since they are related by a simple gauge
transformation. Hence, one may view configuration space as a
circle with circumference \(\pi\).

This is the same for the full gauge-Higgs fields. When a
transition over the sphaleron barrier is considered, the system
starts and ends in the same state (since the different vacua are
related by a large gauge transformation). Nevertheless the winding
number is changed: \(\Delta N_{\rm CS}=1\). The (change in the)
winding number is the physical quantity of interest, since it is
related to the (change in the) baryon number through
(\ref{relCSbar}). The relation between a change in the
Chern-Simons number and \(\Theta\) is (see Appendix A)
\begin{equation}
N_{\rm CS}(t)-N_{\rm
CS}(0)=\frac{2}{\pi}\int_{0}^{t}dt'\dot{\Theta}\sin^2\Theta .
\end{equation}

\begin{figure}[h]
    \centering{\epsfig{figure=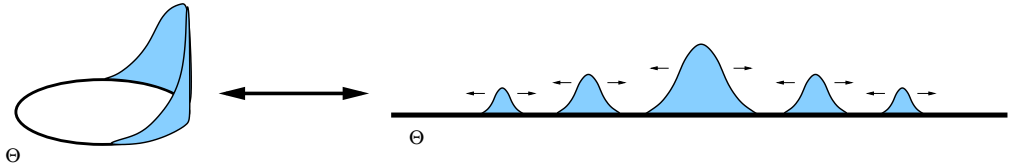,width=0.85\textwidth}}
\caption{The distribution function of $\Theta$ in a periodic potential. On the left it is viewed on the circle $\Theta \in [0,\pi[$, then the distribution function is static. On the right it is unwound to the infinite line $\Theta \in ]-\infty,\infty[ $ accounting thus for the winding number naturally. On the infinite line the system is dynamical and the distribution function diffuses.} \label{figure0}
\end{figure}

Instead of considering the system on a circle and keeping track of
the winding number, it is convenient to "unwind" the system and
consider the system on an infinite line \(\Theta\in
]-\infty,\infty[\) and follow the dynamics of \(\Theta\) (see
figure \ref{figure0}). Thus, we consider the diffusion of
\(\Theta\in ]-\infty,\infty[\), where we keep in mind that
actually we consider the winding number on a circle.

To discuss the diffusion of \(\Theta\) in the next section it is
convenient to derive the Hamiltonian. Firstly, we write the
Langrangian corresponding to (\ref{actiontheta}) and
(\ref{CPactionstheta}) as
\begin{equation}
L=
z_{1}(\Theta)\dot{\Theta}^2+z_2(\Theta)\dot{\Theta}^3-V(\Theta),
\label{LangTheta}\end{equation} with
 \bea
z_1(\Theta)&=&\frac{4\pi }{g^3v}(a_1+a_2\sin^2\Theta),\\
z_2(\Theta)&=&\frac{4\pi v^2}{M^4}[b_1\delta^{1}_{\rm
CP}+b_2\delta^2_{\rm CP}+(b_3\delta^{1}_{\rm
CP}+b_4\delta^{2}_{\rm CP})\sin^2\Theta]\sin^2\Theta. \eea Clearly
the kinetic energy in (\ref{LangTheta}) is unbounded. And, for
instance, a (CP-even) \(\dot{\Theta}^4\) is required to put a
lower bound on the kinetic energy. Such a \(\dot{\Theta}^4\) term
should come from a higher (non-renormalizable) operator, so it is
expected to be suppressed compared to the other CP-even
\(\dot{\Theta}^2\)-term. Therefore the details of the
\(\dot{\Theta}^4\)-term are unimportant for the following
calculations. In fact strictly to first order in \(\delta_{\rm
CP}^1\) and \(\delta_{\rm CP}^2\) (which are assumed to be small),
one can work with the unbounded Lagrangian (\ref{LangTheta}) (and
the unbounded Hamiltonian (\ref{HamTheta})), although it may be
kept in mind that a \(\dot{\Theta}^4\)-term (or higher even powers
of \(\dot{\Theta}\)) are required for a bounded energy. We assume
that these higher-order CP-even terms will make the Lagrangian
convex, so that we can go to a Hamiltonian description. The
conjugate momentum then reads \begin{equation}
p=2z_1(\Theta)\dot{\Theta}+3z_2(\Theta)\dot{\Theta}^2,
\end{equation} and the
Hamiltonian \begin{equation}
H(p,\Theta)=\frac{1}{4z_1(\Theta)}p^2-\frac{z_2(\Theta)}{8[z_1(\Theta)]^3}p^3+V(\Theta).
\label{HamTheta}\end{equation} This is the Hamiltonian that will
be used for the calculation of statistical averages in the next
section.

The point of this section was to show that the CP-odd operators in
\(S^{[8]}\) (\ref{dimeight}) projected onto the one-dimensional
path in phase space gives a \(p^3\)-term in the Hamiltonian and a
\(\dot{\Theta}^3\)-term in the Lagrangian, and to give reasonable
estimates for the coefficients in the projected action and
Hamiltonian. The main subject of this paper, the effect of the
CP-odd terms on the dynamics, will be discussed in the next and
following sections.

\section{Velocity expectation values}\label{sec:avvel}

We study the effect of the action (\ref{dimeight}) on the
diffusion of the Chern-Simons number. For this the Hamiltonian
(\ref{HamTheta}) will be used. We will work to first order in the
coefficients \(\delta_{\rm CP}^1\) and \(\delta_{\rm CP}^2\) (and
hence, to first order in \(z_2\)). So the question is what does
the diffusion of \(\Theta\) look like to first order in
\(\delta_{\rm CP}^1\) and \(\delta_{\rm CP}^2\). In particular,
will the distribution function of \(\Theta\) develop an asymmetry?

The first thing one may notice is that (the magnitude of) the
velocity differs for positive or negative \(\dot{\Theta}\). More
precisely, at a given momentum (\(p\)) or at a given energy
(\(H\)) the magnitude of the velocity depends on its direction.
This implies that the motion over the barrier towards negative
Chern-Simons number differs from the motion towards positive
Chern-Simons number. Hence, an asymmetry is expected to develop.

It is useful to define a quantity that "measures" the asymmetry. A
simple and useful quantity is the mean velocity, at a certain
position, when \(\Theta\) moves either to the right or to the
left. We define \bea v^{\uparrow}(\bar{\Theta})&=& \langle
|\dot{\Theta}|{\cal
H}(\dot{\Theta})\delta(\Theta-\bar{\Theta})\rangle
/\langle {\cal H}(\dot{\Theta})\delta(\Theta-\bar{\Theta})\rangle, \label{velright}\\
v^{\downarrow}(\bar{\Theta})&=& \langle |\dot{\Theta}|{\cal
H}(-\dot{\Theta})\delta(\Theta-\bar{\Theta})\rangle/\langle {\cal
 H}(-\dot{\Theta})\delta(\Theta-\bar{\Theta})\rangle,
\label{velleft}\eea where the brackets denote a thermal average,
and \({\cal H}\) the Heaviside function. When the difference
\begin{equation}\Delta
v(\bar{\Theta})=v^{\uparrow}(\bar{\Theta})-v^{\downarrow}(\bar{\Theta})
\label{veldiff}\end{equation} is non-zero, an asymmetry will
develop. Consider, for instance, the evolution of the distribution
function of \(\Theta\). When the particles\footnote{It is
convenient to think of the distribution function as a normalized
sum of a lot of single particle positions.} that move to the left
do it faster (slower) than the particles that move to the right,
the tail of the distribution function to the left extends more
(less) than to the right. Hence, the distribution function
develops an asymmetry. In particular, we expect expectation values
of odd powers of \(\Theta\) to become negative (positive) when the
distribution function is initially symmetric and thermal.

In the calculation of the velocities (\ref{velright}) and
(\ref{velleft}), firstly one may remark that their numerators are
equal. This follows from the fact that the flux through a point
\begin{equation}\langle\dot{\Theta}\delta(\Theta-\bar{\Theta})\rangle=
\langle|\dot{\Theta}|{\cal
H}(\dot{\Theta})\delta(\Theta-\bar{\Theta})\rangle-
\langle|\dot{\Theta}|{\cal
H}(-\dot{\Theta})\delta(\Theta-\bar{\Theta})\rangle \end{equation}
vanishes. This may be seen from \bea \langle
\dot{\Theta}\delta(\Theta-\bar{\Theta})\rangle &=&
\langle \partial_{p}H\delta(\Theta-\bar{\Theta})\rangle\nn\\
&=&Z_{\pi}^{-1}\int_{\-\infty}^{\infty}dp\int_{-\pi/2}^{\pi/2}d\Theta \partial_p H \delta(\Theta-\bar{\Theta})e^{-\beta H}\nn\\
&=&-Z_{\pi}^{-1}T\int_{\-\infty}^{\infty}dp\int_{-\pi/2}^{\pi/2}d\Theta
\delta(\Theta-\bar{\Theta})\partial_{p}e^{-\beta H}=0.
\label{dotThetais0}\eea Here we used that \(\exp -\beta H\)
vanishes at \(p=\pm \infty\), for which we had to keep in mind
that a \(p^4\)-term in the Hamiltonian is required to make it
bounded and convex. (Alternatively, strictly working to first
order in \(\delta_{\rm CP}^1\) and \(\delta_{\rm CP}^2\), the
momentum integrations in (\ref{dotThetais0}) also vanish.) The
normalization factor \(Z_{\pi}\) is defined as usual, see
(\ref{normfactor}). It will drop out of the calculation of the
velocities \(v^{\uparrow}\) and \(v^{\downarrow}\). The fact that
the flux vanishes, implies that the expectation value of
\(\Theta\) remains constant: \begin{equation}
\langle\Theta(t)-\Theta(0)\rangle=0,
\label{avThetacst}\end{equation} as will be discussed more fully
in the section \ref{sec:remarks}. Hence, asymmetric diffusion will
affect only expectation values of higher odd powers of \(\Theta\).

The result for the numerator on the r.h.s. of (\ref{velright})
(equal to the numerator on the r.h.s. of (\ref{velleft})) is \bea
\langle|\dot{\Theta}|{\cal
H}(\dot{\Theta})\delta(\Theta-\bar{\Theta})\rangle
&=&Z_{\pi}^{-1}\int_{-\infty}^{\infty}dp\int_{-\pi/2}^{\pi/2}d\Theta \partial_p H{\cal H}(p)\delta(\Theta-\bar{\Theta})e^{-\beta H}\nn\\
&=&Z_{\pi}^{-1}Te^{-\beta V(\bar{\Theta})}. \eea The denominator
of (\ref{velright}) is given by \bea \langle {\cal
H}(\dot{\Theta})\delta(\Theta-\bar{\Theta})\rangle&=&
Z_{\pi}^{-1}e^{-\beta V(\bar{\Theta})}\int_{0}^{\infty}dp \,
e^{-\beta\frac{1}{4z_1(\bar{\Theta})}p^2
+\beta\frac{z_2(\bar{\Theta})}{8[z_1(\bar{\Theta})]^3}p^3}\nn\\
&=&Z_{\pi}^{-1}\left[\sqrt{\pi
z_1(\bar{\Theta})T}+T\frac{z_2(\bar{\Theta})}{z_1(\bar{\Theta})}\right]
e^{-\beta V(\bar{\Theta})}, \eea to first order in \(z_2\). The
denominator of (\ref{velleft}) yields \begin{equation}\langle
{\cal H}(-\dot{\Theta})\delta(\Theta-\bar{\Theta})\rangle=
Z_{\pi}^{-1}\left[\sqrt{\pi
z_1(\bar{\Theta})T}-T\frac{z_2(\bar{\Theta})}{z_1(\bar{\Theta})}\right]e^{-\beta
V(\bar{\Theta})}. \end{equation} Hence, up to first order in
\(\delta_{\rm CP}^{1}\) and \(\delta_{\rm CP}^{2}\) we have
\begin{equation}v^{\uparrow
(\downarrow)}(\bar{\Theta})= \sqrt{\frac{T}{\pi
z_{1}(\bar{\Theta})}}
-(+)\frac{T}{\pi[z_{1}(\bar{\Theta})]^{2}}z_2(\bar{\Theta}),
\label{resvel}\end{equation} and \begin{equation}\Delta
v(\bar{\Theta})=
-\frac{2T}{\pi[z_{1}(\bar{\Theta})]^2}z_2(\bar{\Theta}),
\label{resveldiff}\end{equation}

The relative velocity difference, \bea
\Delta v_{\rm rel}(\bar{\Theta})&=&\frac{\Delta v(\bar{\Theta})}{v^{\uparrow}(\bar{\Theta})+v^{\downarrow}(\bar{\Theta})}\nn\\
&=&
-\sqrt{\frac{T}{\pi}}\frac{z_2(\bar{\Theta})}{[z_1(\bar{\Theta})]^{3/2}},
\label{asvelocities}\eea gives a (dimensionless) measure for the
effect of CP-violation on diffusion.

In conclusion, we have established that, when \(\Theta\not = 0\),
the magnitude of the thermal expectation value of the velocity
depends on the direction in which \(\Theta\) moves.

Let us consider what the difference in thermal expectations values
of the velocity implies for the evolution of the distribution
function of \(\Theta\) and $p$, $P(p,\Theta,t)$. When the
momentum, $p$, is thermally distributed independent of the
position \(\Theta\), the velocity difference is present in the
entire ($\Theta$-)space. Now it is not hard to imagine that, when
the particles move faster to the left than to the right, the tail
of an initially symmetric thermal distribution function will
(start to) extend more to the left than to the right. Therefore, a
symmetric thermal distribution function will not remain symmetric.
The diffusion develops asymmetrically when CP-odd terms like
\(\dot{\Theta}^3\) are included in the
lagrangian\footnote{Strictly, the argument given only implies that
\(P(\Theta,p,t)=P(-\Theta,p,t)\) cannot remain to hold as time
evolves. In principle, it might still be possible that \(\int dp
P(\Theta,p,t)= \int dp P(-\Theta,p,t)\) remains true. Since, in
the broken phase at least, we expect the momenta to be thermally
distributed independent of the position, \(\Theta\), we do not
consider this possibility further.}. When we translate this result
back to the gauge-Higgs system it implies that the inclusion of
the CP-odd operators in (\ref{dimeight}) leads to an asymmetric
diffusion of the Chern-Simons number.

\section{Asymmetric diffusion: a conjecture for large times}\label{sec:conj}
In the previous section, it was shown that there is an asymmetry
in the average velocity (\ref{asvelocities}), which implies that
the distribution function will develop an asymmetry. Hence, for
instance, the expectation value of the third power of the
Chern-Simons number, \(\langle [N_{\rm CS}(t)-N_{\rm
CS}(0)]^3\rangle\), will be nonzero. In this section, we will
argue that this expectation value will grow linearly in time in
the broken phase of the SU($2$)-Higgs model.

We assume that after a transition over the sphaleron barrier the
system thermalizes before a following transition. This implies
that the transitions are uncorrelated. This assumption is
reasonable when the height of the barrier, \(E_{\rm sph}\), is
large compared to the temperature, \(T\), and the motion over the
barrier is damped sufficiently. (In our description in terms of
\(\Theta\), this would require \(\Theta\) to be not-too-weakly
coupled to the other modes in the plasma.)

On the basis of this assumption we can establish the
time-dependence of the third power of the Chern-Simons number. We
have established in the previous section that an asymmetry will
develop in the distribution function starting from an symmetric
thermal initial distribution function. Hence, after a time
\(\Delta t\) the expectation value \(\langle [N_{\rm CS}(\Delta
t)-N_{\rm CS}(0)]^3\rangle\) has a nonzero value. During the time
\(\Delta t\) the distribution function has spread over different
vacua. Since we assume that the distribution is thermal in each
vacuum, during the time from \(\Delta t\) to \(2\Delta t\), from
each vacuum the distribution will diffuse asymmetrically in the
same way as it did in the time interval from time \(0\) to time
\(\Delta t\). The thermalization in each vacuum implies that in
each time interval \(\Delta t\), from each vacuum the same
diffusion process takes place (relative to that vacuum). From this
argument follows the result that the expectation value of the
third power of the Chern-Simons number in the presence of CP-odd
dimension-eight operators in the broken phase of the SU($2$)-Higgs
model grows linearly in time. That the repetition of the same
asymmetry in the diffusion process in each time interval leads to
a linear growth, may be seen, for instance, from a random walk
model, as is shown in Appendix B.

In section \ref{subsec:possiblepowers}, we show that in general
(without the above asumption) the expectation value of a third
power of a coordinate $x$, $\langle x^3\rangle$, either stays
constant or grows linearly in time in thermal equilibrium. A
stochastic model that will be introduced later, shows that, with
potential barriers, $\langle x^3\rangle$ grows linearly in time,
and that without barriers $\langle x^3\rangle$ goes to a constant.
This supports the view that barriers are required to ensure that
the asymmetry in the diffusion process is independent of the
position (which vacuum the system is in) and time, from which the
linear behavior follows. In the symmetric phase, different vacua
are not well separated by a barrier, therefore it is for us not
possible to determine the time-dependence of asymmetric
Chern-Simons number diffusion.

To obtain a more detailed conjecture for diffusion in the broken
phase, we extend our one-dimensional model by coupling \(\Theta\)
to the other modes (that form the heat bath at temperature \(T\))
as in the full SU($2$)-Higgs model. In this way the rate for
transitions over the barrier is given by \((gv)^{-3} \Gamma_{\rm
sph}\), with \((gv)^{-3}\) the volume of the sphaleron. Then, we
expect \begin{equation}\langle \Theta^3(t)\rangle\sim
(gv)^{-3}\Gamma_{\rm sph}t \Delta v_{\rm rel, sph},
\label{Theta^3}\end{equation} with the relative velocity
difference (\ref{resvel}) at the sphaleron configuration \(\Delta
v_{\rm rel, sph}=\Delta v_{\rm rel}(\pi/2)\).

By viewing the entire space as made up of blocks of volume
(\(gv)^{-3}\) in each of which a \(\Theta\)-coordinate is
diffusing, one can translate (\ref{Theta^3}) into an expectation
value for the third power of the Chern-Simons number
\begin{equation}\langle
\left[N_{\rm CS}(t)-N_{\rm CS}(0)\right]^3\rangle=c V\Gamma_{\rm
sph}t \Delta v_{\rm rel,sph}, \label{conjCS}\end{equation} with
\(V\) the volume and \(c\) a constant of order one. In going from
(\ref{Theta^3}) to (\ref{conjCS}) we have in a very
non-sophisticated manner included the zero mode for translation
invariance. (It is conventional to include rotational zero modes
in \(\Gamma_{\rm sph}\) and exclude the translational zero modes.)
A better way would have been to include the zero modes in the
parameterization (\ref{gpathparametrization}) and
(\ref{hpathparametrization}). However our aim is to investigate
the occurrence of asymmetries for which the zero modes play no
essential role.

\section{Remarks on statistical averages}\label{sec:remarks}

The Chern-Simons number diffusion is a non-equilibrium process.
Still we, as many others, see e.g. \cite{arnoldmclerran}, use a
thermal (equilibrium) average to calculate the mean or average
velocity of the Chern-Simons number (of \(\Theta\) actually). In
this subsection, we will briefly consider the use of statistical
averages. In the next subsection, we return to dynamical issues
and consider the possible time-dependence that an expectation
value can have.

Consider a particle in one dimension coupled to a heat bath. In
general one expects that a probability distribution function,
\(P(p,x,t)\), of the position, \(x\), and momentum, \(p\), in the
long time limit, goes to a thermal distribution function:
\begin{equation}
P(p,x,t)\stackrel{t\rightarrow\infty}{\rightarrow} Z^{-1}e^{-\beta
H(p,x)}, \end{equation} with \(Z^{-1}\) the normalization factor.
Then, the long-time limit of expectation values can be calculated
using the thermal distribution function. However, in the case that
the Hamiltonian is periodic, the thermal distribution function is
not normalizable. A well-known consequence is that \(\langle
x^2(t)\rangle\) in a flat or periodic potential does not have a
thermal limit-value at large times (since the thermal average is
not defined). Instead it grows linearly in time. Also thermal
averages of other positive powers of \(x\) cannot be calculated.
In particular, since \(\langle x^n\rangle \) is not well-defined
in equilibrium, one cannot conclude from a symmetric potential,
\(V(x)=V(-x)\), that for odd \(n\) it should vanish.

Nevertheless there is still a lot that may be calculated using a
thermal distribution. This is based on the notion that the
long-time limit of the distribution function in a periodic
potential with period \(\pi\) satisfies \begin{equation}\sum_{n}
P(p, x+n\pi, t)
\stackrel{t\rightarrow\infty}{\rightarrow}Z_{\pi}^{-1}e^{-\beta
H(p,x)}, \label{perlimit}\end{equation} where \( Z_{\pi}^{-1}\) is
the normalization factor, with
\begin{equation}
Z_{\pi}=\int_{-\pi/2}^{\pi/2} dx\int_{-\infty}^{\infty} dp \,
e^{-\beta H}. \label{normfactor}
\end{equation}
Equation (\ref{perlimit}) expresses the fact that when the points
\(x+n\pi\) are identified, that is when we view the dynamics of
the system on a circle with circumference \(\pi\), the system
thermalizes as usual. There are two different ways to look at a
periodic system. When the system is considered in terms of the
coordinate \(x\in ]-\infty,\infty[\), the system is dynamical.
There is diffusion, transitions to other classical vacua, etc.
Whereas, when we view the system as living on a circle, with
coordinate \(\phi = x \mbox{ mod}\pi\), the system is static after
thermalization (of course, the dynamics returns when the winding
number is considered).

Thermal averages can be calculated of functions, \(f(p,x)\), that
are either independent of \(x\) or \(\pi\)-periodic in \(x\). In
the long-time limit these functions have a thermal limit value:
\begin{equation}\int_{-\infty}^{\infty}dx\int_{-\infty}^{\infty}dp \; f(p,x)
P(p,x,t) \stackrel{t\rightarrow\infty}{\rightarrow}
Z_{\pi}^{-1}\int_{-\half \pi}^{\half \pi
}dx\int_{-\infty}^{\infty}dp \; f(p,x) e^{-\beta H(p,x)}.
\label{therlim}\end{equation}
 In particular the average velocity can be calculated with
a thermal distribution (as we did in section \ref{sec:avvel}).
Since the average velocity vanishes \begin{equation}\langle
\dot{x}\rangle=0,
\end{equation} one expects that \(\langle x(t)\rangle\) is constant (in the
long-time limit). There exists no such argument for expectation
values of other odd powers of \(x\). For instance, the expectation
value of the time-derivative of \(x^3\), \(\langle
3\dot{x}x^2\rangle\), is itself not well defined. Hence, \(\langle
x^3\rangle\) does not need to vanish and has no thermal
limit-value. Indeed, as we have argued earlier, it may grow in
time.

\subsection{Possible power laws}\label{subsec:possiblepowers}
Next, we will determine on general grounds the possible
time-dependence of \(\langle x^3(t)\rangle\) in a periodic
potential in the long-time limit. We assume that it behaves as a
power law \begin{equation}\langle x^3(t)\rangle = C[P_{\rm in}]
t^p, \label{powlaw}\end{equation} with \(C\) and \(p\) constants.
The brackets denote a classical average, that is, an average over
initial conditions weighted by an initial distribution function
\(P_{\rm in}\). The constant \(C\) may depend on the initial
distribution.

We will use that
 \begin{equation}
 \langle x(t)-x(0) \rangle =0,
 \label{xisconstant}\end{equation}
 see
(\ref{avThetacst}). It is convenient to  choose the origin such
that \(\langle x\rangle \) vanishes at the initial time. Since
then \(\langle x(t)\rangle =0\) for all times, and  the
disconnected contributions to \(\langle x^3(t)\rangle\) vanish.

To determine the possible values for \(p\) it is useful to
consider \begin{equation}\langle [x(t_2)-x(t_1)]^3\rangle=\langle
x^3(t_2)\rangle -3\langle x^2(t_2)x(t_1)\rangle + 3\langle
x(t_2)x^2(t_1)\rangle -\langle x^3(t_1)\rangle.
\label{apppowlaw}\end{equation} When \(t_2-t_1\rightarrow \infty
\), we can apply (\ref{powlaw}) to the left hand side
\begin{equation}\langle
[x(t_2)-x(t_1)]^3\rangle= C[P(t_1)](t_2-t_1)^{p}. \end{equation}
The constant \(C[P(t_1)]\) depends on the distribution function at
time \(t_1\) (instead of the initial distribution function). A
small subtlety is that \(P(t_1)\) should be considered as a
distribution function of the shifted coordinate
\(y(t)=x(t)-x(t_1)\) instead of \(x(t)\).

Consider the correlation functions \(\langle
x(t_2)x^2(t_1)\rangle\) and \(\langle x^2(t_2)x(t_1)\rangle\) on
the right hand side of (\ref{apppowlaw}) in the same limit
\(t_2-t_1\rightarrow \infty\). When the system thermalizes it
follows from the fact that the expectation of \(x\) goes to a
constant (\ref{xisconstant}) that the correlation functions
\(\langle [x(t_2)-x(t_1)]x^2(t_1)\rangle\) and \(\langle
[x(t_2)-x(t_1)]^2 x(t_1)\rangle\) go to a (small) constant in the
limit \(t_2-t_1\rightarrow \infty\). Therefore, when \(p>0\), we
have in this limit
 \bea
\langle x(t_2)x^2(t_1)\rangle &\rightarrow& \langle x^3(t_1)\rangle,\\
\langle x^2(t_2)x(t_1)\rangle &\rightarrow& \langle
x^3(t_1)\rangle.
 \eea
When both \(t_2\) and \(t_1\) are send to infinity also,
(\ref{apppowlaw}) gives
\begin{equation}C[P(t_1\rightarrow \infty)](t_2-t_1)^{p}= C[P_{\rm
in}]t_2^p- C[P_{\rm in}]t_1^p.
\end{equation}
This equation has two possible solutions \bea
p=1,&& \mbox{with } C[P(t_1\rightarrow\infty)]=C[P_{\rm in}]=C,\\
p=0.&& \eea Hence the expectation value \(\langle x^3(t)\rangle\)
either grows linearly in time or stays constant. When the
expectation value \(\langle x^3(t)\rangle\) goes to a constant in
the long-time limit, it is not so that this constant should be
zero. The constant can be non-zero when it depends on the initial
distribution. (In appendix C we consider a stochastic equation
where \(\langle x^3\rangle\) indeed goes to a non-zero constant in
the long-time limit.)

It is reasonable to expect that a similar reasoning can be applied
to connected correlation function of higher powers of \(x\), with
the result that these too are either constant or linearly growing
in the long-time limit. Then, in the case that \(\langle
x^3(t)\rangle\) grows linearly in time, the expectation values of
higher powers of \(x\) are dominated by their disconnected parts:
\begin{equation}\langle x^{2n+1}(t)\rangle \approx
\frac{(2n+1)!}{2^{n-1}3!(n-1)!} \langle x^3(t)\rangle[\langle
x^2(t)\rangle]^{n-1}. \label{long-timedecoherence}\end{equation}
For a linearly growing expectation value of \(x^3\), we get
\(\langle x^{2n+1}\rangle \sim t^n\).

To summarize, on general grounds it has been shown that the
expectation value of \(\langle x^3(t)\rangle\) goes to a constant
or grows linearly in the long-time limit. As we have argued in
section \ref{sec:conj} we expect that, in the broken phase, the
expectation value of the third power of the Chern-Simons number
grows linearly in time. It is encouraging that the result of the
argumentation in section \ref{sec:conj}  is consistent with the
more general analysis presented in this section. Nevertheless, it
is desirable to have a more realistic model that would enable one
to test the conjecture that \(\langle x^3\rangle\) in a periodic
potential with high barriers grows linearly in time. In the next
section we present such a model, namely a simple stochastic
equation.

\section{A stochastic equation}\label{sec:stoch}

In section \ref{to1d} a one-dimensional Lagrangian was derived for
the motion of the Chern-Simons number. In this section we want to
(re)introduce the effect of the (infinite number of) other modes
\cite{khlebnikov}. They provide a heat bath at temperature \(T\).
The simplest way to mimic their effect is by the introduction of a
damping term and a stochastic noise in the equations of motion. In
fact, the motion of the Chern-Simons number in the symmetric phase
is to leading order indeed determined by a stochastic (field)
equation \cite{bodeker}. In the broken phase, the case that is of
interest to us, a local damping term and stochastic force is
probably only suitable for illustrative purposes, and not of
direct quantitative interest for Chern-Simons number diffusion.
Nevertheless the stochastic equation derived below will provide a
simple and realistic model for asymmetric diffusion.

Consider the Lagrangian \begin{equation}
L=\frac{1}{2}\left(\dot{x}^2+\delta\dot{x}^3+d\delta^2\dot{x}^4\right)-V(x),
\label{illLang}\end{equation} with \(\delta\) a small parameter
and \(d\) a dimensionless coefficient. The \(\dot{x}^3\) is the
time-reversal non-invariant term similar (but simpler) as the
\(\dot{\Theta}^3\) term in (\ref{LangTheta}). The \(\dot{x}^4\)
has been added so that the kinetic energy is bounded from below.
We will demand that the Lagrangian is convex, which requires
\(d>9/24\); we choose for the following \(d=1\).

The equation of motion is
\begin{equation}
\ddot{x}\left(1+3\delta\dot{x}+6\delta^2\dot{x}^2\right)=-\partial_x
V.
\end{equation}
We now introduce damping and a stochastic force in the following
way
\begin{equation}
\ddot{x}\left(1+3\delta\dot{x}+6\delta^2\dot{x}^2\right)=-\partial_x
V-\gamma\dot{x}+\xi, \label{stocheq}\end{equation} with \(\gamma\)
the damping coefficient and \(\xi\) a Gaussian white noise \bea
\langle \xi(t)\rangle&=&0,\\
\langle \xi(t)\xi(t')\rangle&=&2\gamma T\delta(t-t'). \eea The
introduction of the damping term (\(\gamma \dot{x}\)) looks
standard. However, there is one subtlety. This may be made clear
by going to the Hamiltonian formulation. The Hamiltonian
corresponding to (\ref{illLang}) (with \(d=1\)) is
\begin{equation}
H=\frac{1}{2}\left(p^2-\delta p^3+\delta^2 p^4\right) +V(x)+{\cal
O}(\delta^3). \end{equation} The Hamilton equations up to order
\(\delta^2\) including noise and damping in the same way as in
(\ref{stocheq}) read \bea
\dot{x}&=&v_p=p-\frac{3}{2}\delta p^2+2\delta^2p^3,\label{illHameq1}\\
\dot{p}&=&-\partial_x V-\gamma v_p+\xi. \label{illHameq2}\eea
However the introduction the damping term as \(\gamma v_p\)
instead of \(\gamma p\) is arbitrary at this point, and requires
an explanation.

One argument for the introduction of the damping term as \(\gamma
v_p\) goes as follows. In a microscopic derivation (for example
using influence-functional techniques \cite{feynman2}), one would
find that integrating out the modes of the heatbath yields a
memory kernel of the form \(\int_{0}^{t}dt'\Sigma(t-t')x(t')\). In
a short-time expansion, this memory kernel will reduce to
\(\gamma\dot{x}\). This reduction is independent of the (possible
complicated) structure of the kinetic term in the Lagrangian or
Hamiltonian. Therefore we expect that in time-reversal
non-invariant theories with complicated kinetic terms, a
derivation from first principles would yield a damping term of the
form \(\gamma \dot{x}\) as introduced in (\ref{stocheq}) and
(\ref{illHameq1}),(\ref{illHameq2}).

Perhaps a more convincing argument is given by considering the
equilibrium distribution function. The equation for the time
evolution of distribution functions, the Fokker-Planck equation,
may be derived in a standard manner, as for instance in
\cite{zinnjustin}. It reads
\begin{equation}
\partial_{t}P=\partial_p\left[\gamma T\partial_p+(\partial_x V)+\gamma v_p\right]P-v_p\partial_x P,
\label{FPeq}\end{equation} with \(P=P(p,x,t)\)  the time-dependent
distribution function. The static solution to this Fokker-Planck
equation is found to be \(\exp-\beta H\). This shows that damping
is correctly introduced in (\ref{stocheq}) and (\ref{illHameq2}).

The remarks of section \ref{sec:remarks} apply to the stochastic
system  (\ref{illHameq1}) and (\ref{illHameq2}). Namely, for
periodic functions or functions that are only dependent on the
momentum \(p\), the thermal average determines the long-time limit
(\ref{therlim})\footnote{Also for periodic functions or functions
that only depend on the momentum \(p\) the ergodic theorem implies
that the statistical average equals the time average. Even when
the time-reversal symmetry is broken \cite{ergodic}.}. For
instance odd powers of the velocity have the following limit
 \begin{equation}\langle
v_{p}^{2n+1}\rangle\rightarrow -\frac{n(2n+2)!}{2^n(n+1)!}\delta
T^{n+1}, \label{velocitylimit} \end{equation} with
\(n=0,1,2,...\). Similarly, it may be concluded that, for example,
the long-time limit of odd powers of \(\sin(2x)\) vanishes when
the potential is symmetric and periodic with period \(\pi\).

The stochastic equation (\ref{stocheq}) introduced here will be
used in appendix \ref{sec:somecalcul} for some sample calculations
with potentials \(V=0\) and \(V=\half \omega^2 x^2\).

\subsection{Numerical simulations}

\begin{figure}[h]
\begin{minipage}[t]{.5\textwidth}
    \centering{\epsfig{figure=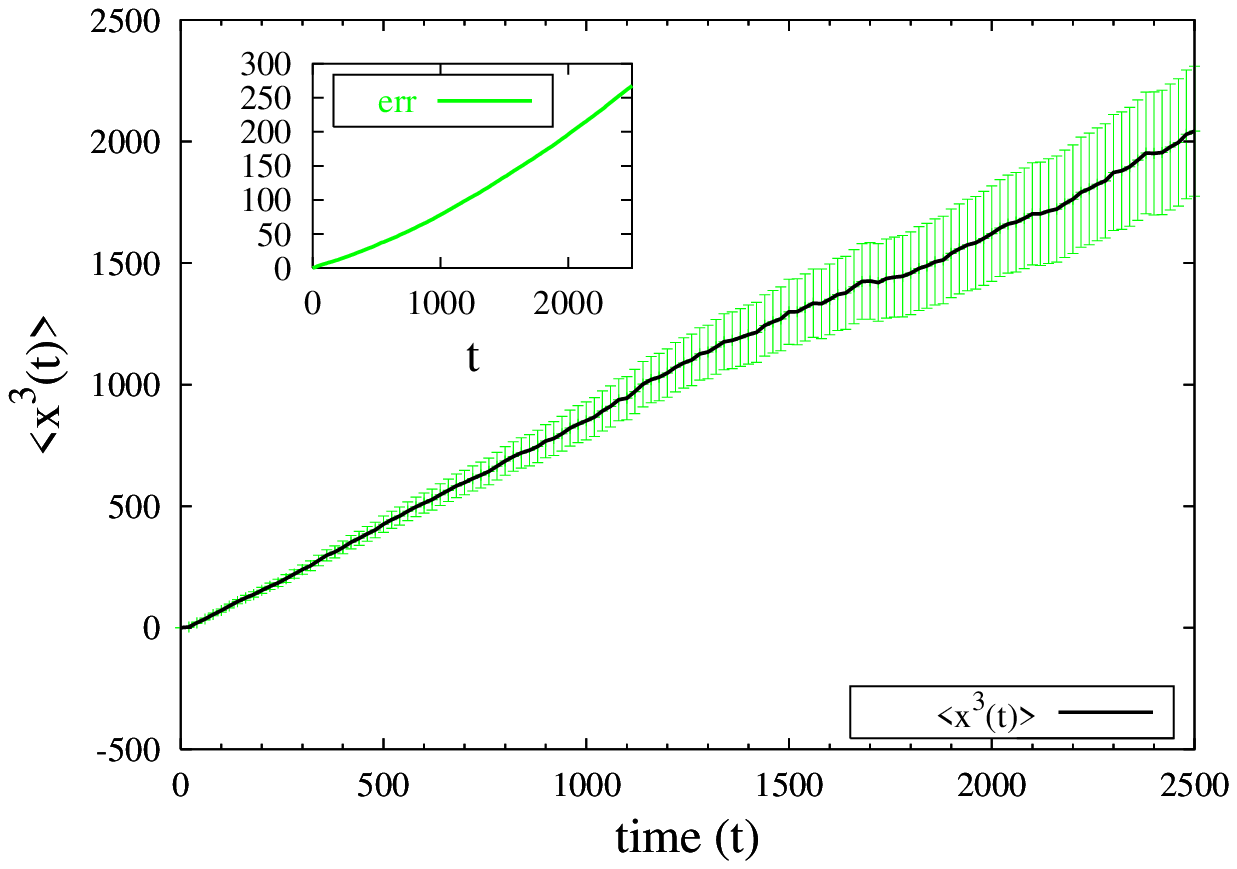,width=0.85\textwidth}}
     \end{minipage}\hfill
\begin{minipage}[t]{.5\textwidth}
    \centering{\epsfig{figure=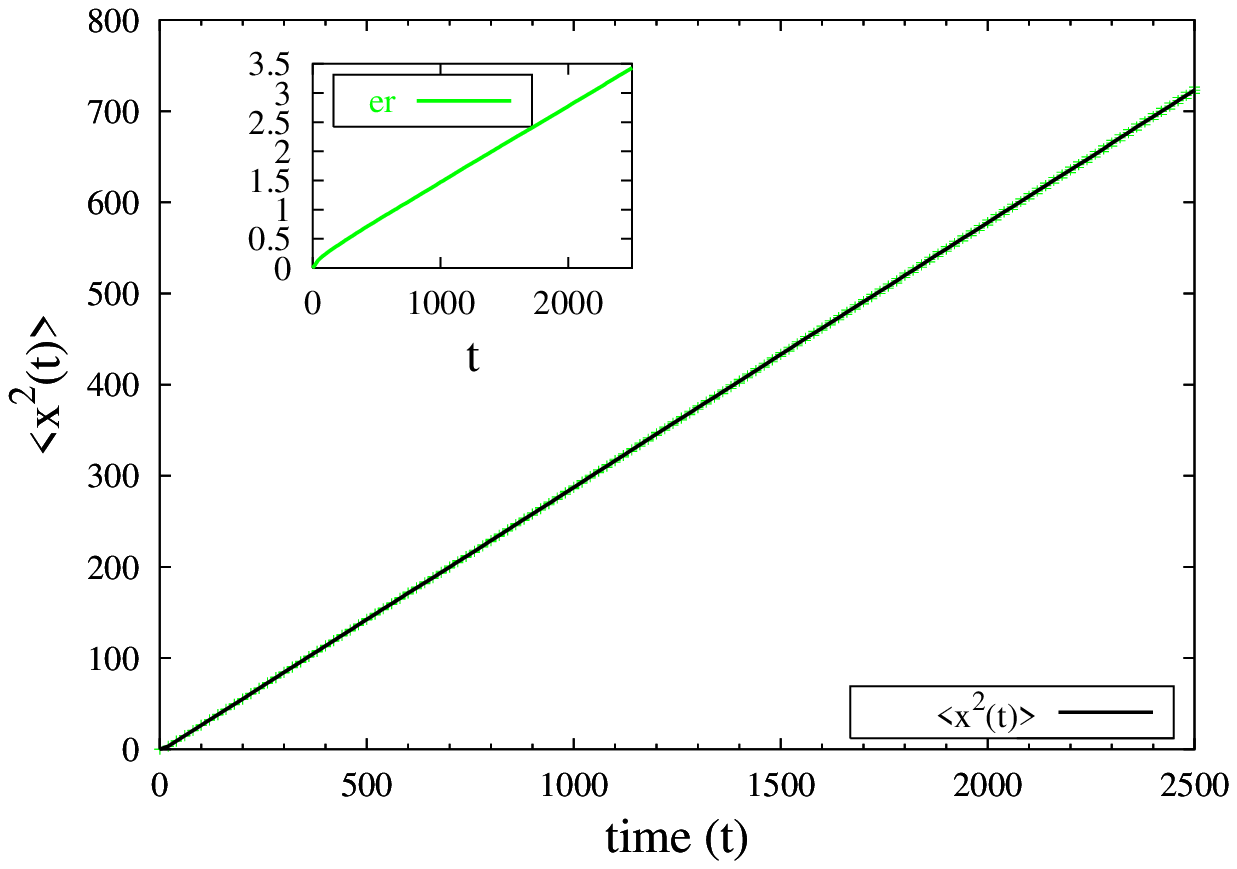,width=0.85\textwidth}}
 \end{minipage}
\caption{\it Evolution of $\langle x^3(t) \rangle$ (left)  and
$\langle x^2(t) \rangle$ (right) for $\delta=0.1$, $\gamma=0.3$
and $V_{\mbox{b}}=4$ (in units where $T=1$) with $8\cdot 10^5$
realizations. Statistical errors of one standard deviation are
also included, together with their time evolution (small
pictures). These grow as $t^{\frac{n}{2}}$ for $\langle x^n(t)
\rangle$. } \label{figure1}
\end{figure}

Perhaps the most relevant case for the stochastic model described
above is when the potential $V$ is a periodic function similar to
(\ref{potentialbarrier}) describing a potential barrier. For
simplicity we take $V$ to be $V(x)=V_{\mbox{\tiny b}}\,
\mbox{sin}^2(x)$ where $V_{\mbox{\tiny b}}$ is then the potential
barrier height. The solution to this case is provided by numerical
analysis of equations (\ref{illHameq1}) and (\ref{illHameq2}) for
a considerable amount of realizations. Average quantities are then
obtained by averaging over all those realizations. The results
from a numerical simulation are displayed in figure \ref{figure1}
for the evolution of the system up to large times ($t \sim 2500$).
One immediately sees from figure \ref{figure1} that $\langle
x^3(t) \rangle$ grows linearly in time. In section
\ref{sec:remarks} we obtained two possible long-time behaviours
for $\langle x^3\rangle$, namely $\langle x^3
\rangle=\mbox{constant}$ or $\langle x^3 \rangle\sim t$. Therefore
the numerical results support our arguments in favour of the
latter for a periodic potential.
 The fact that it grows to a positive value is somewhat surprising since
 our equilibrium considerations showed that the velocity in the negative direction $v^{\downarrow}(\bar{\Theta})$ is larger than the velocity in the positive direction $v^{\uparrow}(\bar{\Theta})$ for $\delta_{\rm CP} > 0$, so $\Delta v=v^{\uparrow}-v^{\downarrow}<0$. This behaviour for the asymmetry $\Delta v$ in the velocities is also found in our numerical simulations. It would seem to follow that $\langle x^3(t) \rangle$ evolves in time to a negative value. This is also supported by the random walk model considerations of appendix \ref{sec:RWM}.{\footnote{Nevertheless
 it is important to say that one cannot make a direct correspondence between the $\Delta$ in the random walk model of appendix \ref{sec:RWM} and the $\delta$ considered here in this stochastic model, not even for the sign.}}
The numerical results however indicate the opposite, which means
that probably non-equilibrium effects determine the sign of growth
of $\langle x^3(t) \rangle$. It might also be that the sign
depends on the details of the interactions, so that a different
model could lead to a different sign.

Expectation values of higher powers of $x$ are dominated by their
disconnected parts in the long time limit as argued in section
\ref{sec:remarks}. Therefore, according to equation
(\ref{long-timedecoherence}) one has
\begin{equation}
f_n(t) \equiv \frac{\langle
x^{2n+1}(t)\rangle}{\frac{(2n+1)!}{2^{n-1}\, 3! (n-1)!}\langle
x^3(t) \rangle \left[ \langle x^2(t) \rangle \right]^{n-1}}
\stackrel{t \rightarrow \infty}{\longrightarrow} 1.
\end{equation}

We have explicitly checked this for $\langle x^5(t) \rangle$ and
$\langle x^7(t) \rangle$ (figure (\ref{figure2}) below).

\begin{figure}[h]
\begin{minipage}[c]{.5\textwidth}
    \centering{\epsfig{figure=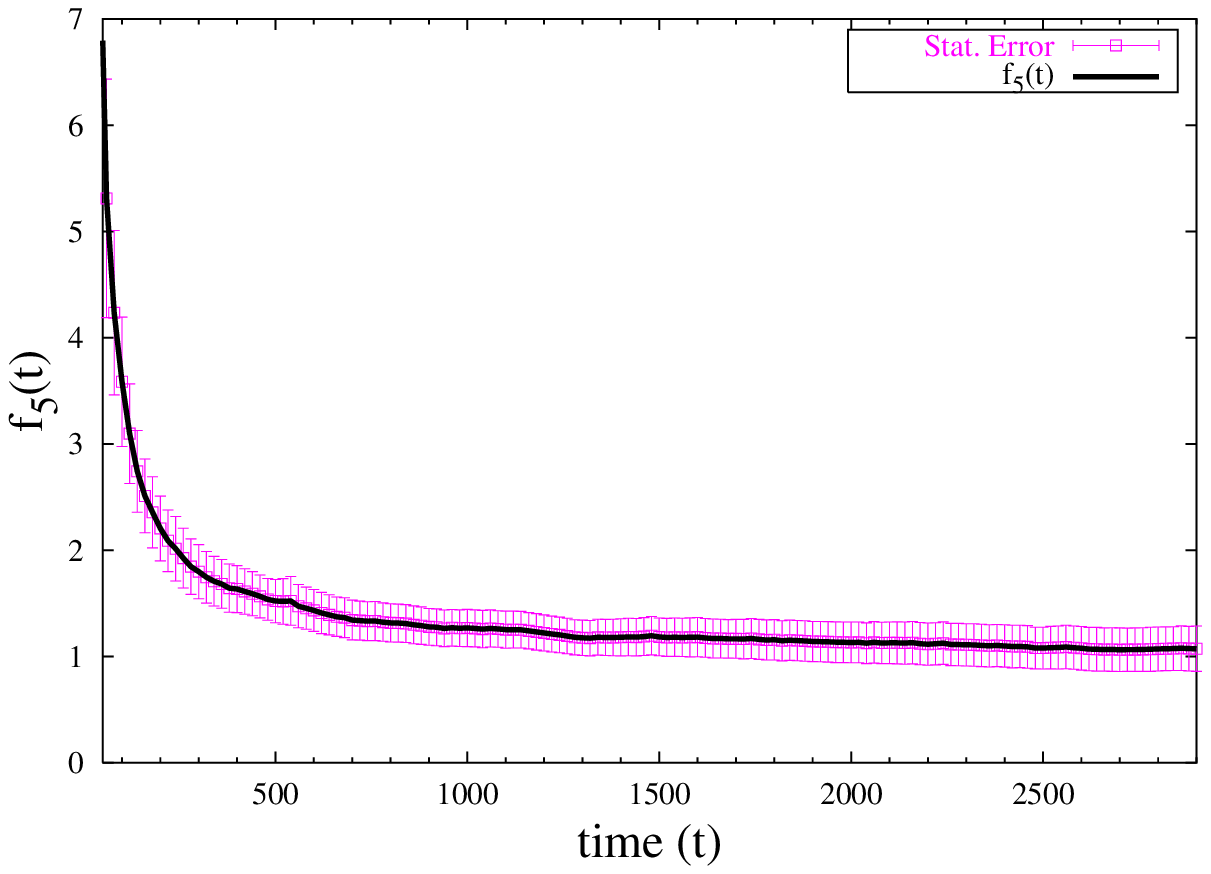,width=.85\textwidth}}
     \label{xfive}
\end{minipage}\hfill
\begin{minipage}[c]{.5\textwidth}
    \centering{\epsfig{figure=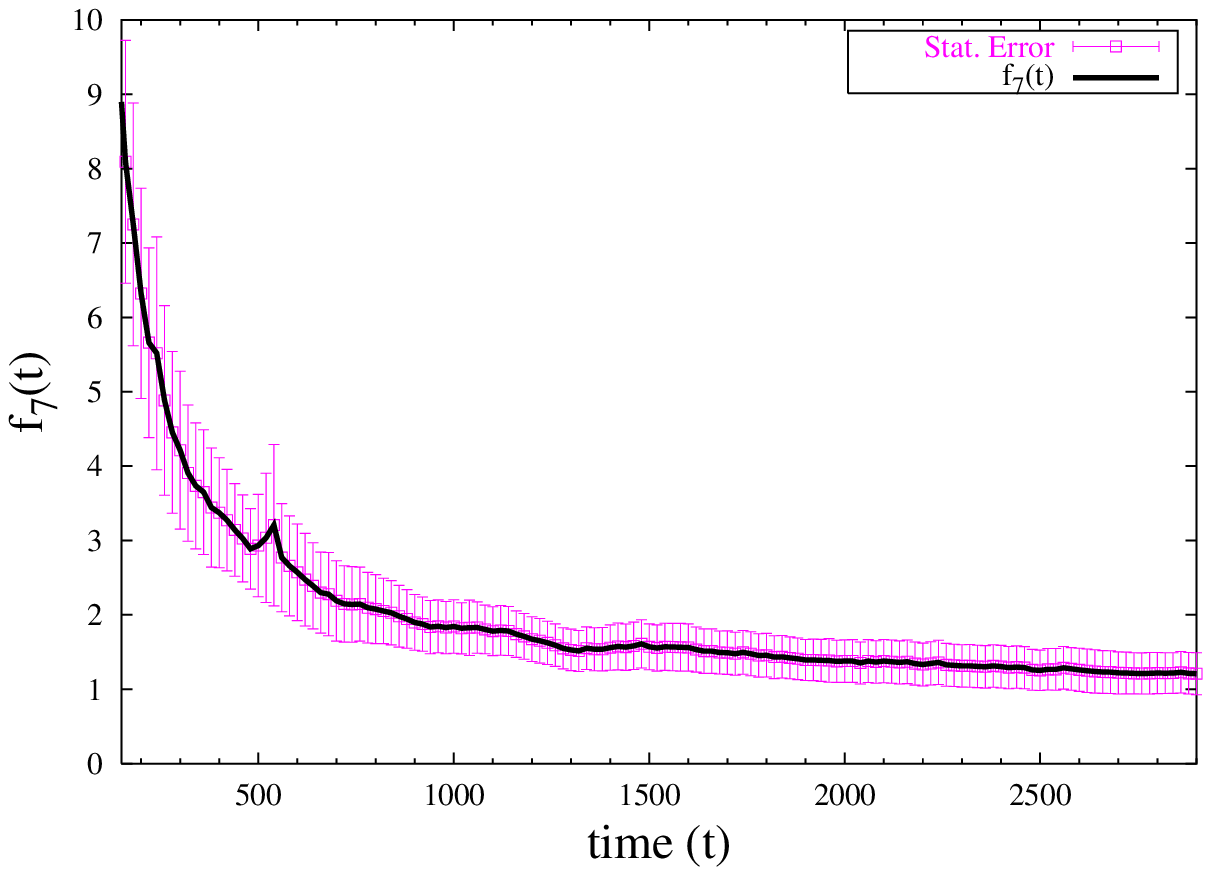,width=.85\textwidth}}
     \label{xseven}
\end{minipage}
\caption{\it Evolution of $f_5(t)=\frac{\langle x^5(t)
\rangle}{10\langle x^3(t) \rangle \langle x^2(t) \rangle}$ (left)
and $f_7(t)=\frac{\langle x^7(t) \rangle}{105\langle x^3(t)
\rangle \langle x^2(t) \rangle^2}$ (right). } \label{figure2}
\end{figure}
In the case of $\langle x^3(t) \rangle$, its disconnected part $\langle x^2(t) \rangle \cdot  \langle x(t) \rangle$ is negligible. This is because $\langle x(t) \rangle$ is constant in time and thus equal to $\langle x(0) \rangle$ which was chosen to be $\langle x(0) \rangle=0$ in our simulations. Actually in the numerical results $|\langle x\rangle| $ is of the order of $0.02\pm 0.1$. The large-time limit of $\langle x^3(t) \rangle$ in figure $\ref{figure1}$ corresponds therefore to its connected part.\\

We have also checked in the simulations that velocity expectation
values $\langle v_{p}^{2n+1}\rangle$ have the limit given by
(\ref{velocitylimit}). Thus we have verified that our stochastic
model thermalizes in the way as expected in section
\ref{sec:remarks}.

\section{On baryon-number generation from asymmetric Chern-Simons number
diffusion}\label{sec:bar}

A natural question to ask is whether the asymmetric diffusion of
the Chern-Simons number may yield a non-zero baryon number in some
out-of-equilibrium situation. This section contains some remarks
on this question. In particular, we will present a simple way in
which asymmetric diffusion of a particle, with coordinate \(x\),
will lead to a (temporary) non-zero expectation value of \(x\)
itself. It is then argued that such a situation occurred during
the electroweak phase-transition yielding a non-zero Chern-Simons
number (and baryon number). In any electroweak baryogenesis
scenario the two important questions are: Is the resulting baryon
asymmetry sufficient to account for the observed asymmetry in the
universe (\ref{obsasymmetry}), and what are the requirements to
prevent a wash out of the generated asymmetry? We will present
 some estimates to answer these questions.

Let us reconsider the diffusion of a particle in one dimension
with coordinate
 \(x\), that starts at \(x=0\). As time evolves the distribution function will
spread. In the long-time limit the distribution will become
Gaussian with on top a small asymmetry (assuming that the
symmetry-breaking terms are small). We have argued that \(\langle
x^3(t)\rangle\) grows linearly in time. To be definite, let us say
it grows in the negative direction. Then the tail of the
distribution in the negative-$x$ direction (-tail) will be larger
than the tail of the distribution function in the positive-$x$
direction (+tail). This difference in the tails accounts for the
negative values of \(\langle x^3\rangle\) (and of expectation
values of higher powers of \(x\)). That the expectation value of
\(x\) remains zero is due to an asymmetry in the distribution
function closer to \(x=0\). (A nice way to imagine this that the
peak of the distribution function (the most probable position)
moves in the positive-$x$ direction. The motion of the peak
would give a growing \(\langle x\rangle\) except that the
contribution of
 the tails is precisely opposite. For higher powers of \(x\) the tails
dominate and a negative and growing value is the result.) With
this picture of the evolution of the distribution function it is
not hard to construct a system for which \(\langle x\rangle\)
itself becomes (temporarily) non-zero. All that one has to do is
to prevent the -tail to compensate for the positive contribution
of the peak moving in the positive-$x$ direction (if the peak does
not move in the positive \(x\)-direction there is still a positive
contribution from the region around \(x=0\) and the argument goes
through unchanged). Consider the diffusion of this particle in a
box with symmetrically placed walls at \(x=\pm a\). The -tail will
hit the wall earlier than the +tail. Then the -tail can no longer
compensate for the asymmetry of the distribution function close to
\(x=0\) (for instance the motion of the peak of the distribution
function in the positive-$x$ direction). Hence, \(\langle x\rangle
\) will grow and become
 non-zero. Eventually, when the system goes to equilibrium,
\(\langle x\rangle \) relaxes back to zero.

Instead of putting the system in a box we could also have let the
system evolve in a harmonic potential \(V=\half\omega^2 x^2\) (to
be superimposed on the periodic potential in which it diffuses).
Basically by the same arguments as above it follows that the
expectation value of \(x\) becomes temporarily non-zero. A
slightly more general case that may be considered, is a system
that is initially in thermal equilibrium in the presence of a
harmonic potential \(V_1=\half\omega_{1}^2 x^2\) which changes at
the initial time to \(V_2=\half\omega_{2}^2 x^2\), with
\(\omega_1^2>\omega_2^2\). In the evolution towards the new
equilibrium state, the expectation value \(\langle x\rangle\) will
temporarily become non-zero. In the next section we will show in a
short-time expansion that \(\langle x\rangle\) will become
non-zero indeed, after such a change in a harmonic potential.

\subsection{Short-time expansion}\label{sec:short-time}
We show in the following that an instantaneous change from a
potential \(V_1=V_{\rm per}+\half \omega_1^2 x^2\) to \(V_2=V_{\rm
per}+\half \omega_2^2 x^2\), with \(V_{\rm per}\) a periodic
potential, will lead to a nonzero expectation value \(\langle
x\rangle\). We will use the Fokker-Planck equation (\ref{FPeq}),
with \(V_2\) as the potential, for the time evolution of the
distribution function \(P(p,x,t)\). The initial distribution
function at time \(t=t_{\rm in}\) is given by
\begin{equation}P_{\rm
in}(p,x)=Z^{-1}e^{-\beta [\half (p^2-\delta p^3+\delta^2
p^4)+V_1(x)]}, \end{equation} and the expectation value \(\langle
x\rangle\) by \begin{equation}\langle x(t)\rangle=\int dx\int dp
\; x \, P(p,x,t). \end{equation} We calculate this expectation
value in a short-time expansion \begin{equation} \langle
x(t)\rangle =\sum_{n}\frac{1}{n!} x_n t^n,
\label{expexpvalx}\end{equation} with coefficients
\begin{equation}x_n=\int dx\int dp \,
x\, \left. \frac{\mbox{d}^n\mbox{ }}{\mbox{d}t^n}
P(p,x,t)\right|_{t=t_{\rm in}}. \end{equation} Define the operator
\begin{equation}
O=\partial_{p}\left[\gamma T\partial_p+\gamma v_p+(\partial_x
V)\right]-v_p \partial_x, \end{equation} such that the
Fokker-Planck equation (\ref{FPeq}) may be written as
\(\partial_{t}P=O P\), with potential \(V=V_2\). The coefficients
\(x_n\) may be calculated by
\begin{equation}x_n=\int dx\int dp \; x \, O^n P_{\rm in}(p,x). \end{equation} The first
three terms in the expansion (\ref{expexpvalx}) vanish \bea
x_0&=&\langle x\rangle_{\rm in}=0,\\
x_1&=&\langle v_p\rangle_{\rm in}=0,\\
x_2&=&\langle (1-3\delta p)\partial_x V_2 -\gamma v_p-3\delta
\gamma (T-p^2)\rangle_{\rm in}=0, \eea where
\begin{equation}
\langle \,..\,\rangle_{\rm in}=\int dx \int dp \,..\, P_{\rm
in}(p,x).
\end{equation}
For \(x_3\) we find
\begin{equation}
x_3=3\delta \langle T\partial^{2}_{x}V_2-(\partial_x
V_2)^2\rangle_{\rm in}.
\end{equation}
When \(V_2=V_1\) this gives \(x_3=0\). To first order in
\(\omega_1^2-\omega_2^2\) the result for \(x_3\) is independent of
the periodic potential, \(V_{\rm per}\),
\begin{equation}
x_3=3\delta (\omega_1^2-\omega_2^2)T.
\end{equation}
This shows that a change in the potential will lead to a
(temporary) non-zero value for \(\langle x\rangle\), even if the
initial and final potential is symmetric.
\subsection{Numerical results}
\begin{figure}[h]
\begin{minipage}[c]{.5\textwidth}
    \centering{\epsfig{figure=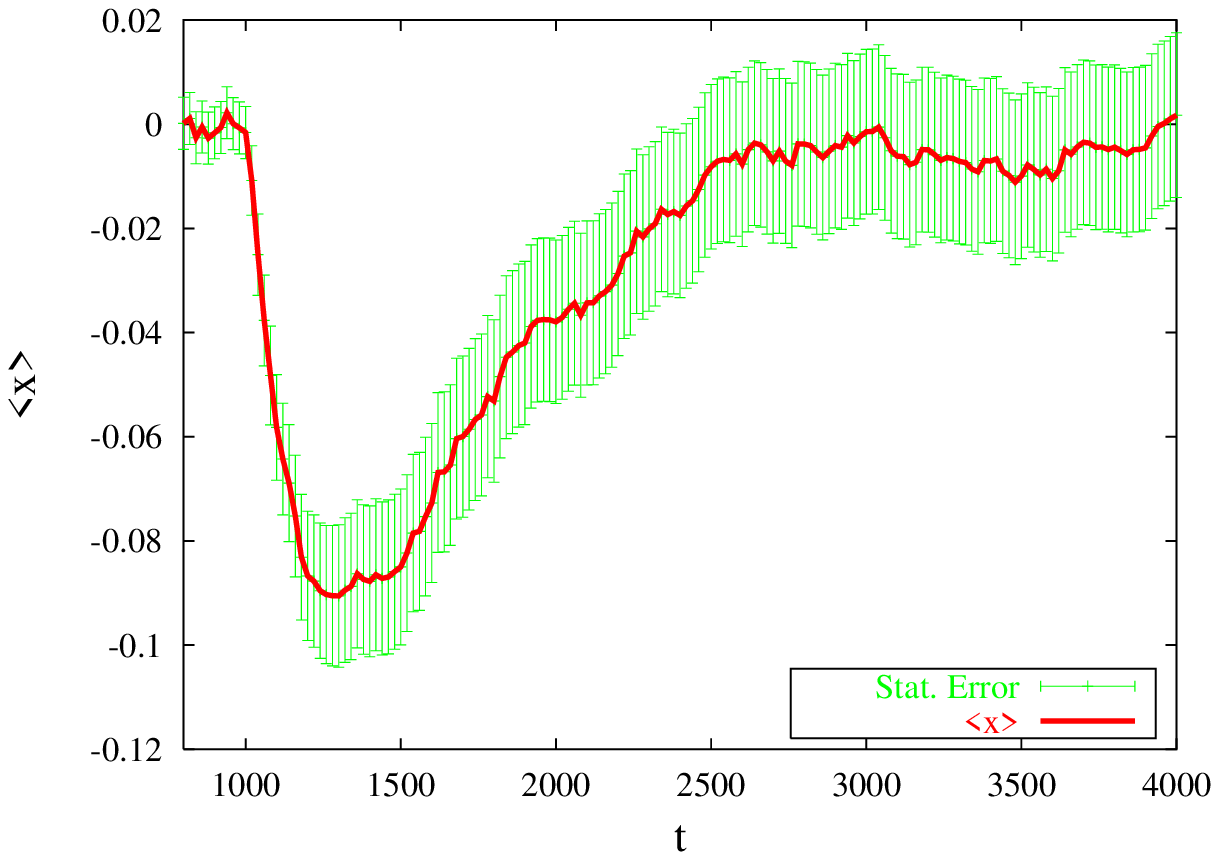,width=.85\textwidth}}
     \label{xquench}
\end{minipage}\hfill
\begin{minipage}[c]{.5\textwidth}
    \centering{\epsfig{figure=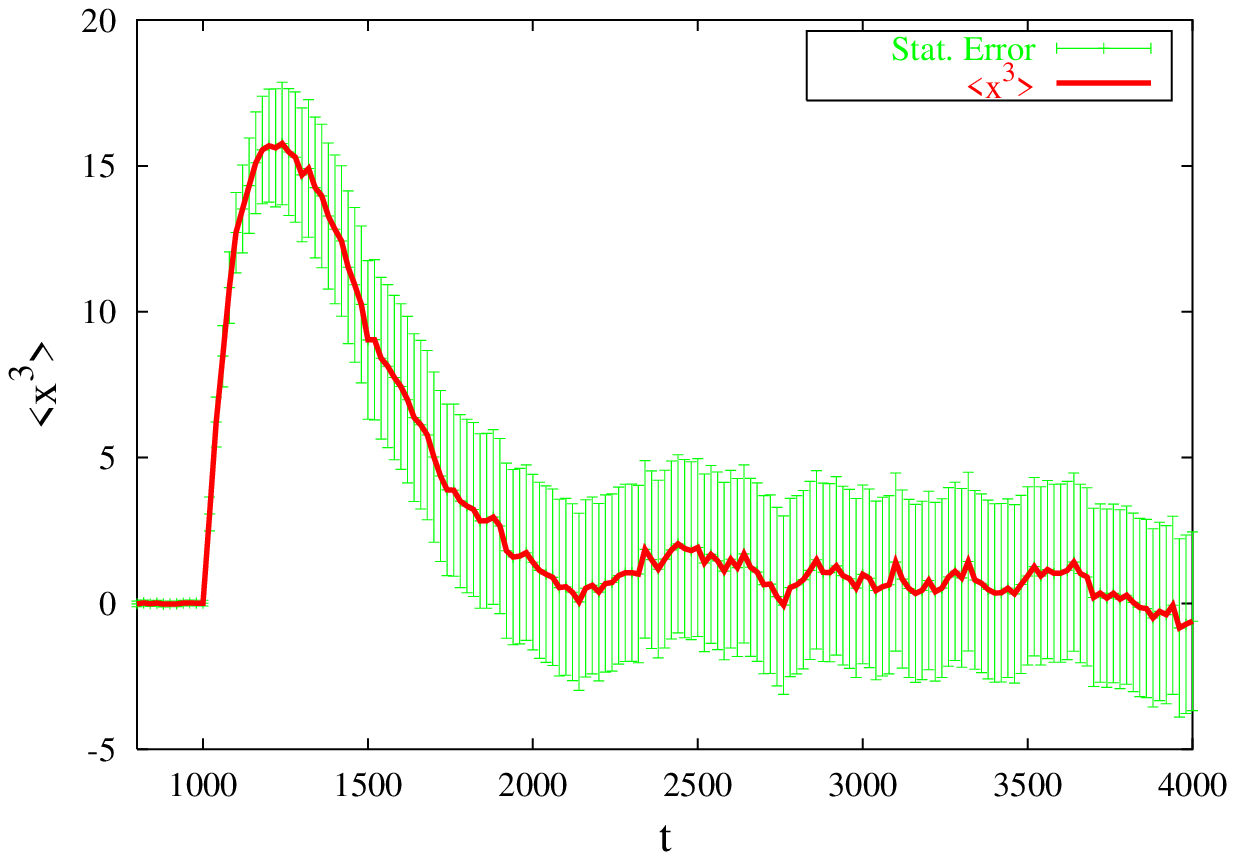,width=.85\textwidth}}
     \label{xcubequench}
\end{minipage}
\caption{\it Evolution of \(\langle x(t)\rangle\) (left) and
\(\langle x^3(t)\rangle\) (right) after the instantaneous
potential change $V_{\rm{before}} \rightarrow V_{\rm{after}}$ (at
$t \sim 1000$) with statistical errors included.}
\label{figurequench}
\end{figure}

To obtain the behaviour of the evolution of \(\langle x\rangle\) for
longer times we numerically solve the stochastic model discussed
in section \ref{sec:stoch}, with the instantaneous potential
change $V_{\rm{before}}=V_{\rm{b}}\sin^2(x)+\half
\omega_{1}^{2}x^2 \longrightarrow
V_{\rm{after}}=V_{\rm{b}}\sin^2(x)+\half \omega_{2}^{2}x^2$ with
$\omega_{1}^{2}$ and $\omega_{2}^{2}$ chosen to be equal to $0.1$
and $0.01$ respectively. The system is first let to be
equilibrated in $V_{\rm{before}}$ and then the instantaneous
potential quench is performed. The numerical results show that
indeed a temporary non-zero value for \(\langle x\rangle\) is
obtained (see figure \ref{figurequench}).

\subsection{Baryon-number generation}

We now turn to the problem of baryon-number generation. The back
reaction of the generated baryons on the Chern-Simons number may
be described by the effective potential given by the free energy
at a given baryon-number \(B\) \cite{arnoldmclerran,khlebnikov}
\bea F(B)=\half \omega_{B}^2 B^2/V, \label{effpot}\eea with
\(\omega_{B}^2=13/6 T^2\) and \(V\) the volume. Due to the
relation between a change in the Chern-Simons number to a change
in the baryon number, the effective potential generates a force on
the Chern-Simons number. As a result a non-zero baryon number will
be "pushed back" to \(B=0\) by the potential. This is the wash out
of baryon number.

In terms of \(\Theta\) the effective potential is given by
\begin{equation}
V_{\rm eff}(\Theta)=\frac{9}{2\pi^2}(gv)^3 \omega_{B}^2\Theta^2,
\end{equation}
where the factor \((gv)^3\) is the inverse volume of a sphaleron.

Now consider the effect of a first-order electroweak
phase-transition. Besides the complicated dynamics of bubble
nucleation and moving bubble walls, the quarks will acquire a mass
through the Higgs mechanism. This produces a small change in the
back-reaction of the quarks on sphaleron transitions. In the
description of this back-reaction in terms of an effective
potential, this means that the potential (\ref{effpot}) is changed
by mass corrections: \(\omega_{B}^2\rightarrow
\omega_{B}^2(1-\sum_i c_i m_i^2/T^2)\), with \(m_i\) the masses of
the different quark species and \(c_i\) dimensionless
coefficients. As argued before, a change in a (harmonic) potential
will lead to a temporary non-zero value of the one-point function.
Hence, the expectation value of the baryon and Chern-Simons number
will acquire temporarily a non-zero value.

Before we discuss how "temporarily" may become forever, we will
turn to the question of how large we may expect the expectation
value of the baryon number to grow. From the sample calculations
in Appendix C, we learn that the asymmetry has typically a maximum
value of \(\sim (1-\omega_2^2/\omega_1^2)\delta\), with \(\delta\)
a dimensionless measure of the amount of CP-violation and
\(\omega_1\) and \(\omega_2\) the two frequencies of the potential
before and after the phase transition. The factor
\((1-\omega_2^2/\omega_1^2)\), that is a measure of the amount of
the departure from equilibrium, is of the order of the mass
correction to the potential (\ref{effpot}). Therefore, it is
dominated by the top quark mass: \((1-\omega_2^2/\omega_1^2)\sim
m_t^2/6T^2\), with \(m_t\) the top quark mass. In the case of
Chern-Simons number diffusion we identify \(\delta\) with the
relative asymmetry in the velocity at the sphaleron configuration
\(\Delta v_{\rm rel,sph}\). Hence, when we consider the motion of
\(\Theta\) in this potential, we estimate for the maximum after
the electroweak phase-transition \begin{equation}\langle \Theta
\rangle_{\rm max}\sim \frac{m_t^2}{6T^2}|\Delta v_{\rm rel,sph}|.
\end{equation}

Imagine the universe made up of blocks of size \((gv)^{-3}\) (the
size of the sphaleron), with in each block a \(\Theta\)-variable.
At the first-order electroweak phase-transition bubbles fill out
the universe and in each block the harmonic potential for
\(\Theta\) changes. The time that the bubbles need to fill out the
universe is much shorter than the time for a single sphaleron
transition \((gv)^{3}\Gamma_{\rm sph}^{-1}\). Therefore, as far as
the motion of \(\Theta\) is concerned, the change in the potential
occurs effectively at the same time in the whole universe. This
implies that the expectation values of \(\Theta\)'s in different
regions in the universe move simultaneously towards their maximum
value. Then we get for the maximum baryon-number density,
\(n_B=\langle B\rangle/V\),
 \begin{equation}
\left.\frac{n_B}{n_{\gamma}}\right|_{\rm max}\approx 4
\frac{(gv)^3}{T^3} \langle \Theta\rangle_{\rm max},
\label{asT=100}\end{equation} with the photon density
\(n_{\gamma}=0.24\mbox{ }T^3\). In this estimate we used that
\(\Theta\)'s in different blocks of space move independently,
similar as in the case for pure diffusion in going from
(\ref{Theta^3}) to (\ref{conjCS}). This is only valid when the
back reaction of the baryons on sphaleron transitions may be
considered as arising from the potential (\ref{effpot}) and
non-localities may be neglected on the length scale \((gv)^{-1}\).
This is what is usually done in the literature, see e.g.
\cite{turok}. If this assumption is incorrect extra suppression
factors may be expected.

When the generated baryon-number is not washed out, the asymmetry
(\ref{asT=100}) results, at the present time, in an asymmetry
\begin{equation}
\left.\frac{n_{B}}{n_{\gamma}}\right|_{\rm max, now} \approx
\left[7\,\delta^1_{\rm CP}+6\, \delta^{2}_{\rm CP}\right]
10^{-4}\;\left(\frac{100 \mbox{ GeV}}{M}\right)^4,
\label{baras}\end{equation} where we used \(g=0.6, m_t=170 \mbox{
GeV}, \mbox{ and } v\sim T\sim 100\mbox{ GeV}\). This rough
estimate for the asymmetry may indicate that in the process
considered here, sufficient baryons may have been generated,
provided that the CP-violation is strong enough and occurs at a
not-too-high energy-scale. It may be noted that the mass \(M\),
that gives the energy scale of new physics, may not be as large as
in scenario's for baryon generation based on CP-violation through
CP-odd dimension-six operators, since (\ref{baras}) is suppressed
by four powers of \(M\) (instead of two). Other constraints for
the scenario based on asymmetric diffusion may be more important
and are discussed in the next section.

\subsection{Constraints}\label{sec:constraints}

The immediate question is whether a once created baryon asymmetry
survive to present times. The decay of the  baryon-number density,
\(n_{B}\), for small densities is given by \cite{arnoldmclerran}
\begin{equation}\frac{\mbox{d} n_{B}}{\mbox{d}t}=-9\beta
\omega_{B}^2\Gamma_{\rm sph}n_{B} \end{equation} The formal
solution is \begin{equation} n_{B}(t)=n_{B}(t_{\rm pt})e^{-R(t)},
\label{linresp}\end{equation} with \begin{equation}
R(t)=9\int_{t_{\rm pt}}^{t}dt'\beta \omega_B^2\Gamma_{\rm
sph}(t'), \end{equation} with \(t_{\rm pt}\) the time of the
electroweak phase-transition and \(\Gamma_{\rm sph}(t)\) the
time-dependent sphaleron rate (its time-dependence enters through
the changing temperature and Higgs expectation value as the
universe cools down). The requirement that a generated baryon
asymmetry survives to present times is
 \begin{equation}R=R(\infty)\lesssim 1.
\label{bound1}\end{equation} This may be translated into a bound
on the phase transition strength \begin{equation}v_{\rm pt}>v_{\rm
cr}\sim 100\mbox{ GeV}, \label{bound2}\end{equation} with \(v_{\rm
pt}\) the Higgs expectation value just after the first-order
electroweak phase transition.

In \cite{nauta} one of the authors believed that the bound to
prevent the wash out of baryon number could be much weaker. This
was based on the realization that that the linear response result
is not directly applicable to the situation where  a non-zero
expectation value of a coordinate is generated by asymmetric
diffusion in a symmetric potential. (Since according to linear
response this expectation value can never become non-zero, which
as was seen in section \ref{sec:short-time}, is not the case.)
However the time scale for equilibration given by linear response
(\ref{linresp}) is still correct. Therefore, the bound
(\ref{bound1}) still applies.

In the scenario discussed there is another bound, since the
asymmetry itself has to be generated after the phase transition.
Therefore, to generate a large asymmetry sufficient sphaleron
transitions must occur after the phase transition. This requires
\begin{equation}
R \gtrsim 1.
\end{equation}
Hence, to generate sufficient baryons without a large wash out in
the scenario that we discuss here, the number of sphaleron
transitions after the phase transition has to lie in a small range
around \(1\). We expect typically something like
\begin{equation}e^{-1}< R<e. \label{bound}\end{equation} This
requirement for the number of sphaleron transitions after the
electroweak phase-transition may be translated into a requirement
on the strength of the phase transition, that may be indicated by
the Higgs expectation value just after the transition, \(v_{\rm
pt}\). In appendix D we show that from (\ref{bound}) it follows
that this Higgs expectation value lies in the range
\begin{equation}v_{\rm
cr}\left(1-\frac{1}{\beta_{\rm pt}E_{\rm sph}}\right) <v_{\rm pt}<
v_{\rm cr}\left(1+\frac{1}{\beta_{\rm pt}E_{\rm sph}}\right),
\label{boundpt}\end{equation} with the inverse inverse temperature
at the phase transition \(\beta_{\rm pt}\sim (100 \mbox{
GeV})^{-1} \). Since, \(\beta_{\rm pt} E_{\rm sph}\sim 45\) the
range of allowed phase transition strengths is quite small. Hence,
for the above scenario for the generation of matter to work, the
strength of the electroweak phase transition has to satisfy strict
bounds (\ref{boundpt}). For a given particle model these bounds
may be translated into bounds for the Higgs mass, as has already
been done for the bound (\ref{bound1}) or (\ref{bound2}) for the
minimal supersymmetric standard model
\cite{quiros},\cite{cline},\cite{lainerummukainen}.

\subsection{Discussion}
Perhaps it is useful to consider the difference between the effect
of the dimension-six operator \(\phi^{\dagger}\phi F\tilde{F}\)
(the lowest-dimensional CP-odd operator) and the dimension-eight
operators \((D\phi)^{\dagger}(D\phi)F\tilde{F}\) and \((FF)
(F\tilde{F})\) in the context of electroweak baryogenesis. As we
have argued the dimension-six operator does not affect the
diffusion of the Chern-Simons number in equilibrium. However in a
non-equilibrium situation it may have an effect. In particular,
when the Higgs expectation value changes in time the dimension-six
operator introduces an effective force \(F_{\rm eff}\sim
\partial_{t}\langle \phi^{\dagger}\phi\rangle\) on the motion of
the Chern-Simons number. This implies that in the case of a
time-dependent Higgs expectation value the diffusion of the
Chern-Simons number yields a non-zero expectation value of
\(N_{\rm CS}\) itself: \(\langle [N_{\rm CS}(t)-N_{\rm
CS}(0)]\rangle\sim \Gamma_{\rm sph} F_{\rm eff}t\). This in
contrast to the asymmetry generated by the dimension-eight
operators in equilibrium that affect only expectation values of
third and higher powers of \(N_{\rm CS}(t)-N_{\rm CS}(0)\).

It is well-known that the dimension-six operator may play a role
in the generation of the baryon asymmetry. For instance, during
the electroweak phase-transition when the Higgs expectation value
grows the effective force can be used to push the Chern-Simons
number and baryon number to some positive value. The problem is
that during the growth of the Higgs expectation value the
sphaleron rate decreases. Therefore it is difficult to generate a
sufficient baryon asymmetry in the short time that the Higgs
expectation grows. A way to circumvent this problem has been given
in \cite{garciabellido}. In that paper the situation was
considered that the reheating temperature after inflation is below
the temperature at which the electroweak phase-transition takes
place. This means that there never was an electroweak
phase-transition. The Higgs expectation value grows during the
reheating of the universe. At this time the CP-odd dimension-six
operator generates the earlier mentioned effective force. Further
it has been argued in \cite{garciabellido} that during the period
of reheating the exponential suppression of the sphaleron rate is
absent. Therefore sufficient sphaleron transitions can take place
and a baryon asymmetry may be generated. Since the sphaleron rate
becomes exponentially suppressed when the plasma thermalizes, a
generated baryon asymmetry may be preserved.

When the asymmetric diffusion of the Chern-Simons number from
CP-odd dimension-eight operators plays a role in the generation of
a baryon asymmetry, as we have argued is possible, then the
time-scales for the generation and the wash out of a baryon
asymmetry are similar. Therefore there is no need to try to avoid
the exponential suppression of the sphaleron rate. The price to
pay is for these similar time scales is that the strength of the
phase transition needs to be finely tuned in order to ensure
sufficient sphaleron transitions after the phase-transition and
avoid a subsequent wash out of baryon number (see section
\ref{sec:constraints}).

Of course, in section \ref{sec:bar} we have neglected various
important dynamical aspects of the problem. Such as the motion of
the bubble walls of a first-order phase transition, the motion of
the baryons and dynamical aspects of their back-reaction on the
motion of the Chern-Simons number, the growth in energy of
sphaleron configuration as a bubble wall passes a certain region
in space (that is as the Higgs expectation value increases), the
non-Brownian beginning of the motion of the Chern-Simons number
\cite{rajantie} etc. These neglected aspects may well modify the
here presented estimate for the final baryon asymmetry. It is even
possible that, being non-equilibrium processes, one of these will
provide, in combination the included CP-violation, a different
mechanism for baryon number generation. If so, the dimension-eight
operators in (\ref{dimeight}) may prove more effective than the
dim-six operator in (\ref{dimsix}) in providing the necessary
CP-violation when the Higgs expectation value is not rapidly
changing. Although all these non-equilibrium phenomena could yield
a non-zero baryon number in combination with CP-violation, we
believe that the mechanism presented in section (\ref{sec:bar}) is
the most natural way to generate an asymmetry on the basis of the
CP-odd dimension-eight operators.

\section{Summary}\label{sec:concl}

We have argued that the inclusion of CP-odd operators, especially
the dimension-eight operators in (\ref{dimeight}), in an effective
action will result in an asymmetric diffusion of the Chern-Simons
number (in the absence of fermions). That is, in thermal
equilibrium the correlation function \(\langle [N_{\rm
CS}(t)-N_{\rm CS}(0)]^3\rangle\) becomes nonzero. In section
\ref{sec:remarks}, we noted that, on general grounds, the
expectation value of the third power of a coordinate either grows
linearly in time or goes to a constant in the long time limit. We
have argued that in the broken phase \(\langle [N_{\rm
CS}(t)-N_{\rm CS}(0)]^3\rangle\) grows linearly in time. A more
detailed conjecture was given in (\ref{conjCS}). Unfortunately,
the study presented here  did not allow us to determine the sign
of the expectation value (the direction of the growth).

Further we noted that asymmetric diffusion may lead to a non-zero
expectation value of the coordinate itself in a non-equilibrium
situation. For instance when the potential is changed (even if
before, during, and after the change the potential is symmetric).
This may have implications for the generation of the baryon
asymmetry in the early universe. Although there are severe constraints on the strength of the electroweak phase-transition, we found that the observed baryon asymmetry could be generated by asymmetric diffusion in combination with the electroweak
phase-transition.

\subsection*{Acknowledgements}
We would like to thank Jan Smit and Jeroen Vink for useful comments and discussions. A.A. was supported by Foundation FOM of the Netherlands.

\appendix

\section{Relation between the Chern-Simons number and \(\Theta\)}

The change in the Chern-Simons number is given by
\begin{equation}
N_{\rm CS}(t)-N_{\rm CS}(0)=\frac{g^2}{32\pi^2} \int_{0}^{t} dt'
\int d^3x F_{\mu\nu}^{a}\tilde{F}^{\mu\nu a}.
\label{appA1}\end{equation} Using the parametrization
(\ref{gpathparametrization}) of the non-contractable loop in
configuration space found by Manton \cite{manton}, equation
(\ref{appA1}) may be rewritten in terms of \(\Theta\):
\begin{equation}
N_{\rm CS}(t)-N_{\rm CS}(0)=
\frac{12}{\pi}\int_{0}^{t}dt'\dot{\Theta}\sin^2\Theta
\int_{0}^{\infty}dr (\partial_{r}f)f(1-f).
\label{appA2}\end{equation} The time and spatial integrations may
be factorized. The spatial integration is completely determined by
the boundary values of the function \(f\) (it does therefore not
depend on the Ansatz that is made for \(f\), given the boundary
conditions)
\begin{equation}
\int_{0}^{\infty}dr (\partial_{r}f)f(1-f)=\frac{1}{6}.
\end{equation}
Hence, the relation between the change in Chern-Simons number and
\(\Theta\) is given by
\begin{equation}
N_{\rm CS}(t)-N_{\rm CS}(0)=
\frac{2}{\pi}\int_{0}^{t}dt'\dot{\Theta}\sin^2\Theta.
\end{equation}

For instance when we change \(\Theta\) from \(\Theta(0)=0\) to
\(\Theta(t)=n\pi\), crossing $n$ barriers, the Chern-Simons
changes as
\begin{equation}
N_{\rm CS}(t)-N_{\rm CS}(0)=n,
\end{equation}
as expected.

\section{Random walk model}\label{sec:RWM}

A random walk model is a convenient toy model to study diffusion.
For asymmetric diffusion we consider a one-dimensional random walk
model with probabilities \begin{equation}P_{ R}=(1+\Delta)/2,\;
P_{L}=(1-\Delta)/2, \label{RWdiffprob}\end{equation} of moving
right ($R$) or left ($L$). The distance the particle moves in one
time step differs also between the two different directions. When
the particle moves to the right it travels a distance \(\Delta
x_{R}\) and when it moves to the left a distance \(\Delta x_{L}\):
\begin{equation}
\Delta x_{R}=(1+\Delta)^{-1},\; \Delta x_{L}=(1-\Delta)^{-1}.
\label{RWdiffdist}\end{equation} The relation
\begin{equation}P_{R}\Delta
x_{R}-P_{L}\Delta x_{L}=0, \label{Rwfluxcons}\end{equation}
implies that the flux vanishes.

Thus, the random walk model defined above contains the two
important features of the motion of the Chern-Simons number
established in section \ref{sec:avvel}. The flux vanishes (at
every point), see (\ref{dotThetais0}),
 and the left and right velocities differ, see (\ref{resveldiff}),
 which translates in the different distances (\ref{RWdiffdist}) travelled
 in one time step in the random walk model.

The distribution function of the number of steps to the right,
\(R\) and the number of steps to the left,\(L\), after a total of
\(N=L+R\) steps, reads
\begin{equation}
P(L,R,N)=\delta_{L+R,N}\frac{N!}{L!R!}\frac{1}{2^{N}}(1-\Delta)^{L}(1+\Delta)^{R}.
\label{RWdistfun}\end{equation} When the particle starts at the
initial time, \(N=0\), at position \(x=0\),
 \(x\) is given by
 \begin{equation}
x=R\Delta x_{R}-L\Delta x_{L}. \label{RWdist}\end{equation} The
relation \(L+R=N\) and (\ref{RWdist}) can be used to convert
(\ref{RWdistfun}) in a distribution function of \(x\). Instead, we
will calculate expectation values. Let us start with \(\langle
x\rangle\). As mentioned before, the vanishing of the flux should
imply that \(\langle x\rangle \) is constant and, since the
particle starts at \(x=0\), is zero. This may be checked for the
random walk model. The expectation value of \(x\) is given by \bea
\langle
x(N)\rangle&=&\sum_{L=0}^{N}\sum_{R=0}^{N}\delta_{L+R,N}\frac{N!}{L!R!}
\frac{1}{2^{N}}(1-\Delta)^{L}(1+\Delta)^{R}\nn\\
&&\times\left[R(1+\Delta)^{-1}-L(1-\Delta)^{-1}\right].
\label{RWexpx}\eea We define \(R'=R-1\) and \(L'=L-1\) and write
(\ref{RWexpx}) as \bea \langle
x(N)\rangle&=&\sum_{L=0}^{N}\sum_{R'=0}^{N-1}\delta_{L+R',N-1}
\frac{N!}{L!R'!}\frac{1}{2^N} (1-\Delta)^L(1+\Delta)^{R'}\nn\\
&-& \sum_{L'=0}^{N-1}\sum_{R=0}^{N}\delta_{L'+R,N-1}
\frac{N!}{L'!R!}\frac{1}{2^N} (1-\Delta)^{L'}(1+\Delta)^{R}.
\label{hm}\eea Since in the first line on the right hand side of
(\ref{hm}) the term \(L=N\) in the sum over \(L\) does not
contribute due to the delta function, and similarly the term
\(R=N\) does not contribute in the second line, the sums in the
first and second line cancel. Hence
\begin{equation}
\langle x(N)\rangle=0,
\end{equation}
in accord with (\ref{dotThetais0}).

Next, consider the expectation value \bea \langle
x^3(N)\rangle&=&\sum_{L=0}^{N}\sum_{R=0}^{N}\delta_{L+R,N}\frac{N!}{L!R!}
\frac{1}{2^{N}}(1-\Delta)^{L}(1+\Delta)^{R}\nn\\
&&\times\left[R(1+\Delta)^{-1}-L(1-\Delta)^{-1}\right]^3.
\label{RWexpx^3}\eea This expression may be evaluated in a similar
manner as (\ref{RWexpx}). The result is \begin{equation}\langle
x^3(N)\rangle=\frac{1}{2}
N\left[(1+\Delta)^{-2}-(1-\Delta)^{-2}\right]. \end{equation}
Hence, for positive \(\Delta\), the expectation value \(\langle
x^3\rangle\) is negative and grows linearly in time.

This calculation supports our argument that when the asymmetry is
independent of space and constant in time the expectation value
\(\langle x^3\rangle\) grows linearly in time.

\section{Some solutions to the stochastic equation}\label{sec:somecalcul}

In this appendix we present some calculations using the Langevin
equation (\ref{stocheq}) with a constant potential or a harmonic
potential.

We start with a calculation of \(\langle x \rangle\) in the case
\(V=0\). The Langevin equation (\ref{stocheq}) with \(V=0\) reads
\begin{equation}
\ddot{x}\left(1+3\delta\dot{x}+6\delta^2\dot{x}^2\right)=-\gamma\dot{x}+\xi,
\end{equation} with \bea
\langle \xi(t)\rangle_{\xi}&=&0,\\
\langle \xi(t)\xi(t')\rangle_{\xi}&=&2\gamma T \delta(t-t'), \eea
and initial conditions \(x(0)=x_{\rm in}\), \(\dot{x}=v_{\rm
in}\).

We solve the equations of motion in an expansion in \(\delta\):
\begin{equation}
x(t)=x_{0}(t)+\delta x_{1}(t)+...,
\end{equation}
where \(x_0\) and \(x_1\) satisfy the equations \bea
\ddot{x}_0&=&-\gamma \dot{x}_0+\xi,\\
\ddot{x}_1&=&-\gamma \dot{x}_1-3\dot{x}_0\ddot{x}_0, \eea with
initial conditions \bea
x_0(0)=x_{\rm in},&& \dot{x}_0(0)=v_{\rm in},\\
x_1(0)=0,&&\dot{x}_1(0)=0. \eea The solution for \(x_0\) reads
\begin{equation}
x_0(t)=x_{\rm in}+\frac{v_{\rm in}}{\gamma}(1-e^{-\gamma
t})+\int_{0}^{t}dt' G(t-t')\xi(t'),
\end{equation}
with the Green function \(G(t-t')=[1-e^{-\gamma (t-t')}]/\gamma\).
The solution for \(x_1\) is
\begin{equation}
x_1(t)=-3\int_{0}^{t}dt'G(t-t')\dot{x}_0(t')\ddot{x}_{0}(t').
\end{equation}
When the average over the stochastic force is performed, we obtain
\bea \langle x_0(t)\rangle_{\xi}&=& x_{\rm in}+\frac{v_{\rm
in}}{\gamma}
(1-e^{-\gamma t}),\label{stochav1}\\
\langle x_1(t)\rangle_{\xi}&=& \frac{3}{\gamma}[v_{\rm in}^2-T]
(\frac{1}{2}-e^{-\gamma t}+e^{-2\gamma t}). \label{stochav2}\eea
We take thermal initial conditions: \(x_{\rm in}=0\) and \(v_{\rm
in}\) is thermally distributed. The average over initial momenta
gives for the initial velocity and velocity squared
\begin{equation}\langle
v_{\rm in}\rangle_{\rm in}=0,\; \langle v_{\rm
in}^2\rangle=T+{\cal O}(\delta). \end{equation} Hence the average
over initial conditions of (\ref{stochav1}) and (\ref{stochav2})
gives \bea
\langle \langle x_{0}(t)\rangle_{\xi}\rangle_{\rm in}&=&0,\\
\langle \langle x_1(t)\rangle_{\xi}\rangle_{\rm in}&=&{\cal
O}(\delta). \eea Therefore the combined thermal and stochastic
average of \(x\) vanishes, \begin{equation}\langle \langle
x(t)\rangle_{\xi}\rangle_{\rm in}=0, \end{equation} up to order
\(\delta\).

This is an explicit check that when the velocity is thermally
distributed, and hence \(\langle \dot{x}\rangle=0\), that then
\(\langle x(t)\rangle =\mbox{constant}\).

The second quantity that we have calculated is \(\langle
x^3(t)\rangle_{\xi}\). We find that it goes to a constant (that
depends on the initial distribution) in the long time limit. For
the initial conditions \(x_{\rm in}=0\) and \(v_{\rm in}\)
thermally distributed, we find \begin{equation}\langle \langle
x^3(t)\rangle_{\xi}\rangle_{\rm in}\rightarrow
-\frac{21}{2\gamma}\delta \left(\frac{T}{\gamma}\right)^2.
\end{equation}
Hence, differently than we have argued for the diffusion of the
Chern-Simons number, in this case the expectation value of the
third power does not grow linearly in time. Since there are no
high barriers, we don't expect the velocity to thermally
distributed, independent of the position in space. Therefore the
asymmetry in the velocities does not need to be independent of the
position in space. In this way the argument for the linear growth
of the expectation of \(x^3(t)\) does not hold in this case.
Hence, the fact that \(\langle x^3(t)\rangle\) goes to a constant
when \(V=0\), does not contradict the expected linear growth of
\(\langle [N_{\rm CS}(t)-N_{\rm CS}(0)]^3\rangle\) in the long
time limit. It does raise the following question. When we add a
periodic potential \(V_{\rm b}\sin^2(x)\), how does the long time
behavior of \(\langle x^3(t)\rangle\) depend on \(V_{\rm b}\)? We
know that for \(V_{\rm b}=0\) it goes to a constant, and we have
argued that for \(V_{\rm b}/T>>1\) it grows linearly in time. So,
at what value of \(V_{\rm b}\) does the behavior change?

Next, we consider the time-evolution in a harmonic potential
\(V=\half \omega_{2}^2 x^2\). As initial distribution we use for
the position \(\exp-\beta \half \omega_1^2 x_{\rm in}^2\) and
\(v_{\rm in}\) is thermally distributed. This situation may be
viewed as follows. The system is in thermal equilibrium before the
initial time, \(t=0\), with a potential \(V=\half \omega_1^2
x^2\). At the initial time the harmonic potential changes \(\half
\omega_1^2 x^2\rightarrow\half\omega_2^2 x^2\). We calculate the
following time evolution of \(\langle x\rangle\). In the case
without damping, \(\gamma =0\), We get to first order in
\(\delta\)
\begin{equation}
\langle x(t)\rangle = \frac{\delta T}{\omega_2}
\left(1-\frac{\omega_2^2}{\omega_1^2}\right) \sin(\omega_2
t)[1-\cos(\omega_2 t)]. \label{expxnodamp}\end{equation} For the
case \(\gamma>>\omega\), we present only the contribution of the
slowest decaying mode
\begin{equation}
\langle x(t)\rangle =- \frac{3\delta T\omega_2^2}{2\gamma^3}
\left(1-\frac{\omega_2^2}{\omega_1^2}\right)e^{-(\omega_2^2/\gamma)t}
+{\cal O}(e^{-2(\omega_2^2/\gamma)t}).
\label{expxdamp}\end{equation} We learn from (\ref{expxnodamp})
and (\ref{expxdamp}) that the expectation value \(\langle
x\rangle\) becomes non-zero after the potential is changed, as we
have argued in section \ref{sec:bar}. And further that the
magnitude of \(\langle x\rangle\) is proportional to
\((1-\omega_2^2/\omega_1^2)\) and \(\delta\).

\section{Bound on the strength of the phase transition}
In this appendix we show how the bound (\ref{bound})
\begin{equation} e^{-1}<R<e, \label{appbound}\end{equation} with
\begin{equation}R=9\int_{t_{\rm
pt}}^{\infty}dt' \beta \omega_{B}^2\Gamma(t'),
\label{R}\end{equation} may be translated into a bound on the
Higgs expectation immediately after the electroweak phase
transition, \(v_{\rm pt}\).

In order to simplify the following calculations we take \(\kappa\)
in (\ref{sphrate}) to be constant, then
 \begin{equation}9\beta
\omega_{B}^2 \Gamma_{\rm sph}=\frac{39}{2} \kappa T e^{-\beta
E_{\rm sph}}. \end{equation} As a further simplifying
approximation, we will assume that the Higgs expectation value
stays constant after the phase transition and that the only time
dependence enters through the temperature as \cite{kolbturner}
\begin{equation}T^2=0.03 \frac{M_{\rm
pl}}{t}, \end{equation} where \(M_{\rm pl}=1.2 \mbox{ }10^{19}\)
GeV is the Planck mass.

When the Higgs expectation value, \(v\), is time-independent so is
the sphaleron energy \(E_{\rm sph}\), and the integration in
(\ref{R}) can easily be performed. The result is
 \begin{equation}R=K\frac{M_{\rm
pl}}{E_{\rm sph}}e^{-\beta_{\rm pt}E_{\rm sph}},
\label{resR}\end{equation} with \(K=1.08\; \kappa\) and
\(\beta_{\rm pt}\) the inverse temperature at the electroweak
phase-transition.

It is useful to determine the value  \(v_{\rm pt}=v_{\rm cr}\) for
which \begin{equation}R=1. \end{equation} Inserting the result
(\ref{resR}) for \(R\) and taking the natural logarithm gives
 \begin{equation}
\ln R=\ln K\frac{M_{\rm pl}}{E_{\rm sph}}-\beta_{\rm pt}E_{\rm
sph}=0. \label{expR}\end{equation} From \cite{arnoldmclerran} we
obtain the value for \(\kappa=1.3\mbox{ }10^4\). Inserting this
value for \(\kappa\) in \(K\) and then in (\ref{expR}) yields
\begin{equation}\beta_{\rm
pt}E_{\rm sph}=45. \end{equation} This gives for the phase
transition strength
 \begin{equation}v_{\rm cr}\sim T_{\rm
pt}\sim 100 \mbox{ GeV}. \end{equation} In \cite{moore} the time-
or temperature-dependence of the Higgs expectation value is taken
into account in the derivation of \(v_{\rm cr}\). We are more
interested in the window of phase-transition strengths determined
by (\ref{appbound}).

A small shift of \(v_{\rm pt}\) by an amount of \(\delta v\),
\(v_{\rm pt}=v_{\rm cr}+\delta v\), saturates the lower bound in
(\ref{appbound}) \begin{equation}R(v_{\rm cr}+\delta v)=e^{-1}.
\end{equation} This
equation gives for \(\delta v\)
  \begin{equation}\frac{\delta
v}{v_{\rm cr}}(1-\beta_{\rm pt}E_{\rm sph})=-1. \end{equation}
Thus the bound (\ref{appbound}) translates into the following
bound for the phase-transition strength \(v_{\rm pt}\)
\begin{equation}v_{\rm
cr}\left(1-\frac{1}{\beta_{\rm pt}E_{\rm sph}}\right) <v_{\rm pt}<
v_{\rm cr}\left(1+\frac{1}{\beta_{\rm pt}E_{\rm sph}}\right),
\end{equation}
for \(\beta_{\rm pt}E_{\rm sph}>>1\).

\end{document}